\documentclass[a4paper,11pt]{article}
\pdfoutput=1
\usepackage{jcappub}

\usepackage{amsmath}
\usepackage{mathtools}
\usepackage{amssymb}
\usepackage{graphicx}
\usepackage{color}
\usepackage{enumerate}
\usepackage[colorlinks=true,citecolor=blue,urlcolor=blue]{hyperref}
\usepackage{soul}
\usepackage[version=4]{mhchem}
\usepackage{tabularx}

\graphicspath{ {Figures/} }

\newcommand{\Eq}[1]{Eq.~\eqref{#1}}

\title{Indirect Detection of Dark Matter Absorption in the Galactic Center}

\author[a]{Kimberly K.~Boddy,}
\author[b]{Bhaskar Dutta,}
\author[b]{Addy J.~Evans,}
\author[b]{Wei-Chih Huang,}
\author[a]{Stacie Moltner,}
\author[b]{Louis E.~Strigari}

\affiliation[a]{Texas Center for Cosmology and Astroparticle Physics, Weinberg Institute, Department of Physics, The University of Texas at Austin, Austin, TX 78712, USA}

\affiliation[b]{Department of Physics and Astronomy, Mitchell Institute for Fundamental Physics and Astronomy, Texas A\&M University, College Station, TX 77843, USA}

\abstract{We consider the nuclear absorption of dark matter as an alternative to the typical indirect detection search channels of dark matter decay or annihilation.
In this scenario, an atomic nucleus transitions to an excited state by absorbing a pseudoscalar dark matter particle and promptly emits a photon as it transitions back to its ground state.
The nuclear excitation of carbon and oxygen in the Galactic Center would produce a discrete photon spectrum in the $\mathcal{O}(10)$ MeV range that could be detected by gamma-ray telescopes.
Using the \texttt{BIGSTICK} large-scale shell-model code, we calculate the excitation energies of carbon and oxygen.
We constrain the dark matter-nucleus coupling for current COMPTEL data, and provide projections for future experiments AMEGO-X, e-ASTROGAM, and GRAMS for dark matter masses from $\sim$ 10 to 30 MeV. We find the excitation process to be very sensitive to the dark matter mass and find that the future experiments considered would improve constraints on the dark matter-nucleus coupling within an order of magnitude.}

\begin{document}

\maketitle


\section{Introduction}
\label{sec:intro}

Astrophysical systems are an excellent source for investigating dark matter (DM) interactions with the Standard Model (SM).
Traditional indirect detection searches rely on observing the visible SM particles produced from DM annihilation or decay.
As a consequence, DM-dominated systems such as dwarf spheroidal galaxies or the Galactic Center (GC) of the Milky Way provide ideal targets for such searches.

If DM is asymmetric~\cite{Kaplan:2009ag} and stable, the standard annihilation and decay channels are not necessarily accessible to produce an indirect detection signal.
There is, however, a broader theoretical landscape of DM models in which observable signals arise from DM processes beyond the standard paradigm~\cite{Boddy:2022knd}.
For example, DM could be absorbed by (the nucleus of) an atom, resulting in (nuclear) atomic excitation and subsequent relaxation via emission of a monochromatic photon~\cite{Vergados:2021ejk,Pospelov:2008jk}.
Absorption of DM can also occur in semiconductors, resulting in multiphonon excitation~\cite{Hochberg:2016sqx}.
In these works, the absorption of DM was considered in the context of direct detection and neutrino experiments.

In this work, we study a novel indirect detection signal arising from DM absorption by nuclei, resulting in nuclear line emissions that cannot be generated from known astrophysical processes.
For example, line emissions generated from collisional excitations between interstellar gas/dust and cosmic rays, gamma ray emission due to neutron capture, or inverse Compton scattering~\cite{Benhabiles-Mezhoud_2013,2021ExA....51.1225D, Manna:2023vsv} occur below $\sim$ 10 MeV~\cite{1998PASP..110..637D, 2004A&A...413..817I, Wang:2007sa,2021ExA....51.1225D}.
Since the absorption process involves both a DM particle and baryon, the high DM density and high baryon density environment of the GC is ideal for our indirect search.

The molecular gas in the inner Galaxy is rich in molecular hydrogen, helium, and carbon monoxide (CO)~\cite{2016MNRAS.456.3596G}.
Of these components, CO is the most apt to study, because its abundance in the Galaxy is well-quantified through molecular transitions. 
Therefore, we focus on the nuclear excitation states of \ce{^{12}C} and \ce{^{16}O}, which we calculate using the \texttt{BIGSTICK} large-scale shell-model code~\cite{johnson2018bigstick,Johnson:2013bna}.
The excitation energies are on the order of 10 to 100 MeV, but we limit our scope to a maximum of $\sim$ 30 MeV, corresponding to the highest observed excitation state~\cite{KELLEY201771,TILLEY19931}.

We use COMPTEL data~\cite{1993ApJS...86..657S} to constrain our model for DM masses from $\sim$ 10 to 30 MeV, and we provide sensitivity projections for future experiments.
COMPTEL observed gamma rays with energies from $\sim$ 1 to 30 MeV~\cite{Schoenfelder:2000bu} and is the only existing gamma-ray data in the energy range of interest.
Proposed future gamma ray telescopes such as AMEGO-X~\cite{Fleischhack:2023ube, Caputo:2022xpx}, 
e-ASTROGAM~\cite{e-ASTROGAM:2018jlu}, and GRAMS~\cite{GRAMS:2021tax} would improve the sensitivity in the MeV regime.
Therefore, we also provide sensitivity estimates for DM absorption with these experiments.

The organization of this paper is as follows.
In Section \ref{sec:nuclear} we describe the model of nuclear excitation via absorption of a pseudoscalar DM particle.
In Section \ref{sec:distFxs} we provide the velocity distribution functions and density profiles we use to model the DM halo and CO gas.
In Section \ref{sec:flux} we show the calculation of the expected gamma ray flux from the nuclear de-excitation of \ce{^{12}C} and \ce{^{16}O}.
In Section \ref{sec:constraints} we present and discuss our results for coupling constraints from COMPTEL data, as well as projections for future MeV gamma-ray experiments.
We conclude in Section \ref{sec:conclusions}.


\section{Nuclear Absorption}
\label{sec:nuclear}

The nuclear absorption process can work for various types of DM particle, e.g., scalar, vector, pseudoscalar, axial vector, and fermion~\cite{Pospelov:2008jk,Dror:2019dib,Dror:2019onn}. 
Here, we consider the nuclear excitation of a nucleus $n$ through the absorption of a pseudoscalar DM particle $\chi$, 
\begin{equation}
    n + \chi \rightarrow n^*,
\end{equation}
with interaction Lagrangian
\begin{equation}
    \label{eq:lag}
    \mathcal{L} = g_\chi (\partial_\mu \chi) \bar{N} \gamma^\mu \gamma^5 N,
\end{equation}
where $N$ is nucleon and $g_\chi$ is the DM-nucleon coupling constant (with dimension ${\rm energy}^{-1}$).
The same bilinear nucleon term is also needed for axial vector DM interactions, while the $\bar{N} \gamma^\mu N$ and $\bar{N} N$ terms are needed for vector and scalar DM candidates.
The DM particle type also could be scalar or vector. The parameter space sensitivity (described in section 5) will not be significantly affected due to the type of DM particle.
In this work, we do not consider any particular origin for DM, which can be model-dependent.
All of the DM might not consist of this few MeV-DM candidate we are considering.
In this work, we do not consider any particular origin for DM, which can be model-dependent.
The origin of this DM can be in the early or the late universe.
In the late universe, the DM can emerge from primordial black holes ($10^{16}$ g - $10^{17}$ g), decay of heavy particles into our light DM candidate, etc.~\cite{Dent:2024yje, Agashe:2022jgk, Coogan:2020tuf, Xie:2024eug,Cheek:2021odj,Baer:2014eja}.
If the light DM is produced in the early universe by a thermal or a non-thermal mechanism, the stability needs to be guaranteed in the model.
However, if it is produced in the late universe, the lifetime is not an issue.
Moreover, the couplings between DM and any SM particle are model dependent (e.g., see Ref.~\cite{DiLuzio:2020wdo}), and we only need the DM effective couplings to nucleons in this work.
We assume that this pseudoscalar DM populates all of the DM in the Galactic Center.

The absorption cross section in the center-of-mass frame is
\begin{equation}
   \label{eq:abs_xsec_CM}
    \sigma^\text{abs}_{n \chi}(v_\text{rel}) = \frac{\pi}{6(2J_n+1)} g_\chi^2 g_A^2 S^\text{GT}
    \frac{|\vec{p}_\chi|^2}{v_\text{rel}} \frac{m_n + \Delta E}{E_\chi E_n} \delta(E_\chi + E_n - m_n - \Delta E),
\end{equation}
where $J_n$ is the nuclear spin, $g_A=1.27$ is the axial form factor~\cite{PDG2000}, $\vec{p}_\chi$ is the DM momentum, $v_\text{rel} \equiv \left| \vec{v}_n - \vec{v}_\chi \right|$ is the relative velocity between the DM particle and the nucleus, $E_\chi$ is the incident DM energy, $E_n$ is the initial energy of the nucleus, $m_n$ is mass of the nucleus, $\Delta E$ is the nuclear excitation energy, and 
\begin{equation}
\label{eq:GT_strength}
    S^\text{GT}=|\langle \Phi_f|| \sum_{i=1}^A \frac{1}{2}\hat{\sigma_i} \hat{\tau_0}|| \Phi_i \rangle|^2
\end{equation}  
is the Gamow–Teller (GT) transition strength between the nucleus excited state $\Phi_f$ and nucleus ground state $\Phi_i$.
Ref.~\cite{Dutta:2022tav} shows that in the long-wavelength limit in which the transfer momentum approaches zero, GT transitions are the dominant contribution to the total cross section among the multipole operators.
Therefore, we approximate the DM absorption cross section by considering only the GT transition.
For other types of DM (e.g., scalar, vector, axial vector), the GT transition might not dominate.
Depending on the spin and the parity of the DM and the nucleus, the dominant operator may be one of the following: electric monopole $\hat{\rho}$, electric dipole $\hat{d}$, magnetic moment $\hat{\mu}$, isovector spin dipole $\hat{D^\sigma}$, or spin $\hat{\sigma}$, which is equivalent to the GT operator.

\begin{figure}[t]
    \centering
    \includegraphics[width=0.8\textwidth]{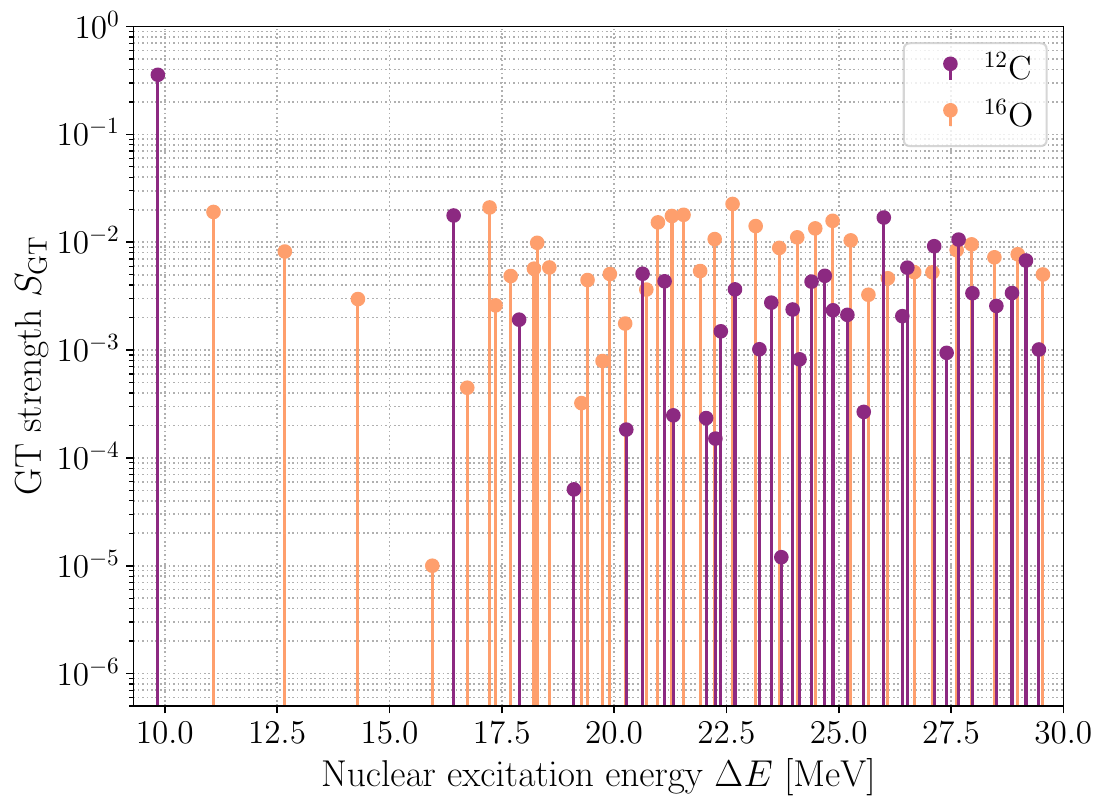}
    \caption{GT strengths for \ce{^{12}C} and \ce{^{16}O}, calculated using large-scale shell-model code \texttt{BIGSTICK}. Each line represents one excited state of the nucleus. Since de-excitation occurs directly to the ground state, the emitted photon energy is equal to the excitation energy $\Delta E$.}
    \label{fig:GT_strengths}
\end{figure}

We use the large-scale nuclear shell-model code \texttt{BIGSTICK}~\cite{johnson2018bigstick,Johnson:2013bna} to calculate the nuclear excitation energy and corresponding GT transition strength in \Eq{eq:GT_strength} of each ground-to-excited state transition for \ce{^{12}C} and \ce{^{16}O}.
In \texttt{BIGSTICK}, we employ the YSOX interaction~\cite{PhysRevC.46.923,Yuan:2012zz}, which is exactly diagonalized within the full $psd$ shell-model space ($0p_{1/2}$, $0p_{3/2}$, $0d_{5/2}$, $1d_{1/2}$, $0d_{3/2}$). The calculated GT strengths for each transition are shown in Fig.~\ref{fig:GT_strengths}.

Nuclear excitation energies and strengths have also been measured in nuclear experiments.
For example, the NDS database~\cite{NDS} collects nuclear data such as energy, transition type, and transition strength.
Nevertheless, there are only a few GT transitions with documented energies and strengths, so the shell-model calculation is necessary in this study.

\texttt{BIGSTICK} is able to compute and produce excitation energies as high as $\mathcal{O}(100)$ MeV.
However, for light nuclei such as \ce{^{12}C} and \ce{^{16}O}, the observed maximal excitation energy is $\sim30$ MeV~\cite{KELLEY201771,TILLEY19931}.
Thus, we do not consider nuclear excitations above $\sim30$ MeV.


\section{Galactic Distribution Functions}
\label{sec:distFxs}

\begin{figure}
    \centering
    \includegraphics[width=0.8\textwidth]{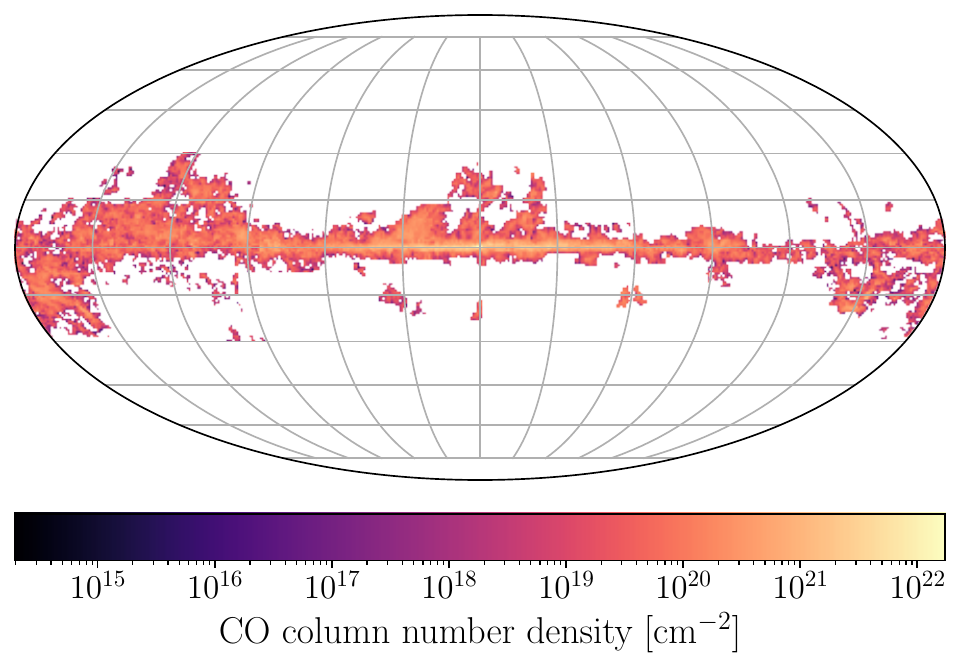}
    \caption{All-sky map of the column number density of CO used in our analysis. We assume the CO number density traces that of H$_2$, and calculate the H$_2$ number density from CO intensity maps provided in Ref.~\cite{Johannesson:2018bit}.}
    \label{fig:densityMap}
\end{figure}

We assume a Navarro-Frenk-White (NFW) density profile for the Milky Way DM halo,
\begin{equation}\label{eq:rho_DM}
    \rho(r) = \frac{\rho_s}{(r/r_s) \left[ 1 + (r/r_s) \right]^{2}},
\end{equation}
where $r$ is the radius from the GC, and the density and radial scales are fixed at $\rho_s = 2.0 ~ \text{GeV/}\text{cm}^3$ and $r_s = 8.1 ~ \text{kpc}$, respectively~\cite{2019MNRAS.487.5679L}.
The CO gas density distribution is modeled using data from \texttt{GALPROP} v57\footnote{https://galprop.stanford.edu/}~\cite{Porter:2017vaa, Porter:2021tlr} for galactic longitude $0.25^\circ \leq l \leq 359.75^\circ$ and latitude $|b| \leq 89.75^\circ$, each with $0.5^\circ$ bins.
To calculate the CO mass density, we start with the CO intensity map (in units of K km s$^{-1}$) used in Ref.~\cite{Johannesson:2018bit} and originally measured by Ref.~\cite{2001ApJ...547..792D}.
To convert from intensity to number density, we use the mass conversion factor from CO to molecular hydrogen (H$_2$), $X_{\text{CO}} = 2 \times 10^{20}$ cm$^{-2}$ (K km s$^{-1}$)$^{-1}$~\cite{Johannesson:2018bit}.
This then gives us the number density of H$_2$, which we assume is traced by the CO number density.
The total gas mass of H$_2$ (CO) is given by Ref.~\cite{Johannesson:2018bit} as $0.67 \times 10^9 M_\odot$ ($0.74 \times 10^9 M_\odot$). This provides us with a normalization factor to determine the CO mass density from the H$_2$ mass density, averaged per radial bin. Note that this averaging process may induce a systematic error in the determination of the number density of the gas. Figure~\ref{fig:densityMap} shows the all-sky map of the CO column number density used for our analysis.
In practice, we use the CO mass density for the remainder of our calculations.

For the DM and nuclei $n \in \{ \ce{^{12}C}, \ce{^{16}O} \}$ velocities, we assume Maxwell-Boltzmann distributions of the form
\begin{align}
    f_\chi(r, \vec{v}_\chi) &= \frac{\rho_\chi(r)}{(2\pi)^{3/2} (\sigma_\chi(r))^3} \text{e}^{-\vec{v}_\chi^2 / 2(\sigma_\chi(r))^2}, \\
    f_n(r, \vec{v}_n) &= \frac{\rho_n(r)}{(2\pi)^{3/2} (\sigma_n(r))^3} \text{e}^{-(\vec{v}_n - \vec{v}_\text{circ}(r))^2 / 2(\sigma_n(r))^2},
\end{align}
where $\vec{v}_\text{circ}(r)$ is the circular baryon velocity, taken to be the rotation curve fit for $R_\odot = 8.5~\text{kpc}$ from Ref.~\cite{Clemens1985} (see Eq.~(5) and Table~3).
We note that more recent estimates find $R_\odot \approx 8.28~\text{kpc}$~\cite{2009ApJ...707L.114G}.

The 1D velocity dispersions we use, in km/s, are given by
\begin{align}
    \sigma_\chi(r) &\simeq \frac{1}{\sqrt{3}}\left( -43 \ln{(r/r_s)} + 288 \right), \label{eq:DM_disp}\\
    \sigma_n(r) &\simeq
    \begin{cases}
        -2.0 \ln{(r/r_s)} + 3.3 & r \leq r_s \\
        3.3 & r > r_s.
    \end{cases} \label{eq:CO_disp}
\end{align}
We obtain the 1D DM velocity dispersion (which is related to the 3D velocity dispersion by $\sigma^\text{1D} = \sigma^\text{3D}/\sqrt{3}$) from a single fit to all of the 3D DM velocity dispersion profiles from the FIRE simulation results of Ref.~\cite{2022MNRAS.513...55M} for $r \lesssim 50 ~ \text{kpc}$.
We fit the baryon velocity dispersion $\sigma_n(r)$ to the 1D CO-bright H$_2$ velocity dispersion of Ref.~\cite{2017A&A...607A.106M}. 
Since the baryon velocity dispersion from Ref.~\cite{2017A&A...607A.106M} only includes data to $r \sim 8.2 ~ \text{kpc}$, we assume a constant $\sigma_n$ for $r>r_s$.

We normalize the velocity distribution functions such that
\begin{align}
    \int_0^{v_\text{esc}(r)} d^3 v_\chi ~ f_\chi(r,\vec{v}_\chi) &= \rho_\chi(r), \label{eq:dm_norm}\\
    \int_0^{v_\text{esc}(r)} d^3 v_n ~ f_n(r,\vec{v}_n) &= \rho_n(r), \label{eq:b_norm}
\end{align}
where the escape velocity of the Milky Way is taken to be $v_\text{esc}(r) = \sqrt{2} v_\text{circ}(r)$. 
We note that this relation underestimates the escape velocity compared to recent estimates based on Gaia DR2~\cite{2021A&A...649A.136K, 2018A&A...616L...9M}, Gaia DR3~\cite{Roche:2024gcl}, and the RAVE survey~\cite{Piffl:2013mla}.


\section{Photon flux}
\label{sec:flux}
For a DM particle absorbed by nucleus $n$, the photon spectrum from the de-excitation of the nucleus is
\begin{equation}
\label{eq:photon_spectrum}
    \frac{dN^{n \chi}}{dE_\gamma} = \delta(\Delta E - E_\gamma),
\end{equation}
where $E_\gamma$ is the true photon energy. 
The differential flux or intensity (in units of cm$^{-2}$ s$^{-1}$ sr$^{-1}$ MeV$^{-1}$) of the resulting de-excitation photon is
\begin{equation}
\label{eq:diff_flux}
    \frac{d^2 \Phi}{dE_\gamma d\Omega} = \frac{1}{4\pi}
    \frac{dN^{n \chi}}{dE_\gamma} \int ds 
    \int d^3v_n \frac{f_n(r(s, \psi), \vec{v}_n)}{m_n} 
    \int d^3v_\chi \frac{f_\chi(r(s, \psi), \vec{v}_\chi)}{m_\chi} 
    \sigma^\text{abs}_{n \chi}(v_\text{rel}) v_\text{rel} ,
\end{equation}
where $s$ is the line-of-sight distance, $\psi$ is the angle between the line of sight and the line from the point of observation (taken to be our Sun) to the center of the Milky Way halo, and $m_\chi$ is the DM mass.

Working in the non-relativistic limit $(p_\chi = m_\chi v_\chi)$, we write the delta function in \Eq{eq:abs_xsec_CM} in terms of $v_\chi$ to obtain
\begin{equation}
   \label{eq:abs_xsec_CM-3}
    \sigma^\text{abs}_{n \chi}(v_\text{rel}) = \frac{\pi}{6(2J_n+1)} g_\chi^2 g_A^2 S^\text{GT}
    \frac{|\vec{p}_\chi|^2}{v_\text{rel}} \times \frac{m_n + \Delta E}{E_\chi E_n} \frac{\delta(v_\chi - v_\chi^+)}{m_\chi v_\chi^+},
\end{equation}
where
\begin{equation}\label{eq:v_chi_plus}
    v_\chi^+ = \sqrt{\frac{2(\Delta E - m_\chi - \frac{1}{2}m_n v_n^2)}{m_\chi}}.
\end{equation}
Since the numerator under the square root must be positive and $v_\chi^+ < v_\text{esc}$, the bounds on the baryon velocity distribution integral in \Eq{eq:b_norm} become 
\begin{equation}
    \sqrt{ \frac{2(\Delta E - m_\chi - \frac{1}{2}m_\chi v_\text{esc}^2)}{m_n}} \leq v_n \leq \sqrt{\frac{2(\Delta E - m_\chi)}{m_n}}.
\end{equation}
The kinematics of the nuclear excitation process are extremely sensitive to $m_\chi$ such that to achieve a nonzero photon flux, the values of $\Delta E$ and $m_\chi$ must be very close to one another.
For DM velocity $v_\chi^+$ bounded by $v_\text{esc} \approx 10^{-3}$, the difference between $\Delta E$ and $m_\chi$ is maximized when $v_n = 0$, for which Eq.~\eqref{eq:v_chi_plus} gives $(\Delta E/m_\chi - 1) \lesssim 5\times10^{-7}$.

The line-of-sight integral is from the location of the detector at $s = 0$ to the source location at
\begin{equation}
    s = \sqrt{R_\text{halo}^2 - R_\odot^2\sin^2\psi} + R_\odot\cos\psi,
\end{equation}
where $R_\text{halo} = 50 ~ \text{kpc}$ is the maximum galactocentric radius considered.
The galactocentric radial coordinate $r$ is related to the line-of-sight distance $s$ by 
\begin{equation}
    r(s, \psi) = \sqrt{R_\odot^2 + s^2 - 2 R_\odot s\cos\psi},
\end{equation}
where $\psi$ is related to galactic longitude and latitude by $\cos\psi = \cos{l} \cos{b}$.


\section{Current constraints and future Projections}
\label{sec:constraints}

To determine constraints on the DM-nucleus coupling $g_\chi$, we numerically calculate the theoretical differential flux in \Eq{eq:diff_flux} for each nuclear excitation energy $\Delta E$ of $\ce{^{12}C}$ and $\ce{^{16}O}$ and for a range of DM masses.
For each $\Delta E$, there is only a narrow range of DM masses $m_\chi~\in~(E_\gamma - 10^{-2} ~\text{MeV},~E_\gamma)$ that give a non-zero differential flux, due to the kinematics of the interaction requiring that
\begin{equation}
    \Delta E = m_\chi + \frac{1}{2}m_\chi v_\chi^2 + \frac{1}{2}m_n v_n^2.
\end{equation}

To approximate the observed spectrum at the detector, the true photon spectrum in \Eq{eq:photon_spectrum} is convolved with a Gaussian distribution.
For an experiment with energy resolution $\epsilon E_\gamma$, the probability of observing a photon of energy $E$ from a gamma ray of true energy $E_\gamma$ is \cite{Perelstein:2010at}
\begin{equation}
    R_\epsilon(E-E_\gamma) \approx \frac{1}{\sqrt{2\pi\epsilon^2 E_\gamma^2}} \exp \left[-\frac{(E-E_\gamma)^2}{2 \epsilon^2 E_\gamma^2} \right].
    \label{eq:gaussian}
\end{equation}
The energy resolutions for the experiments we consider are listed in Table~\ref{tab:expt_parameters}.


\subsection{Constraints from COMPTEL data}
\label{sec:constraints-COMPTEL}

Intensity maps for the inner galactic plane were calculated by Ref.~\cite{Strong:1998ck} using COMPTEL observations for 1--3 MeV, 3--10 MeV, and 10--30 MeV energy bins over the angular range $330.0^\circ \leq l \leq 359.0^\circ, 0.5^\circ \leq l \leq 30.0^\circ$ and $|b| \leq 5^\circ$.
We fit the upper and lower limits of the $E^2 \times \text{intensity}$ COMPTEL results to a piecewise linear function over each energy bin (see Fig.~3 of Ref.~\cite{Strong:1998ck}) and integrate the fitted function over each energy bin.
We use this information to compare with the prediction from DM absorption: after we integrate over the same angular range as the COMPTEL data and divide by solid angle, we convolve the differential flux in \Eq{eq:diff_flux} with \Eq{eq:gaussian}. 
We then multiply by $E^2$ and integrate over the relevant energy bin, using an energy resolution of $\epsilon = 2\%$.
While Ref.~\cite{Strong:1998ck} used binning of $1^\circ$ for $l$ and $5^\circ$ for $b$, we use $0.25^\circ$ bins for both $l$ and $b$.
Since our baryon density data ranges from $0.25^\circ \leq l \leq 359.75^\circ$ and $|b| \leq 89.75^\circ$ in $0.5^\circ$ bins, using $0.25^\circ$ bins allows us to match the same angular range, as well as use all CO density data points in the range we consider.
For intermediate values of $l$ and $b$ we use linear interpolation to determine the CO density.

\begin{figure}
    \centering
    \includegraphics[width=0.8\textwidth]{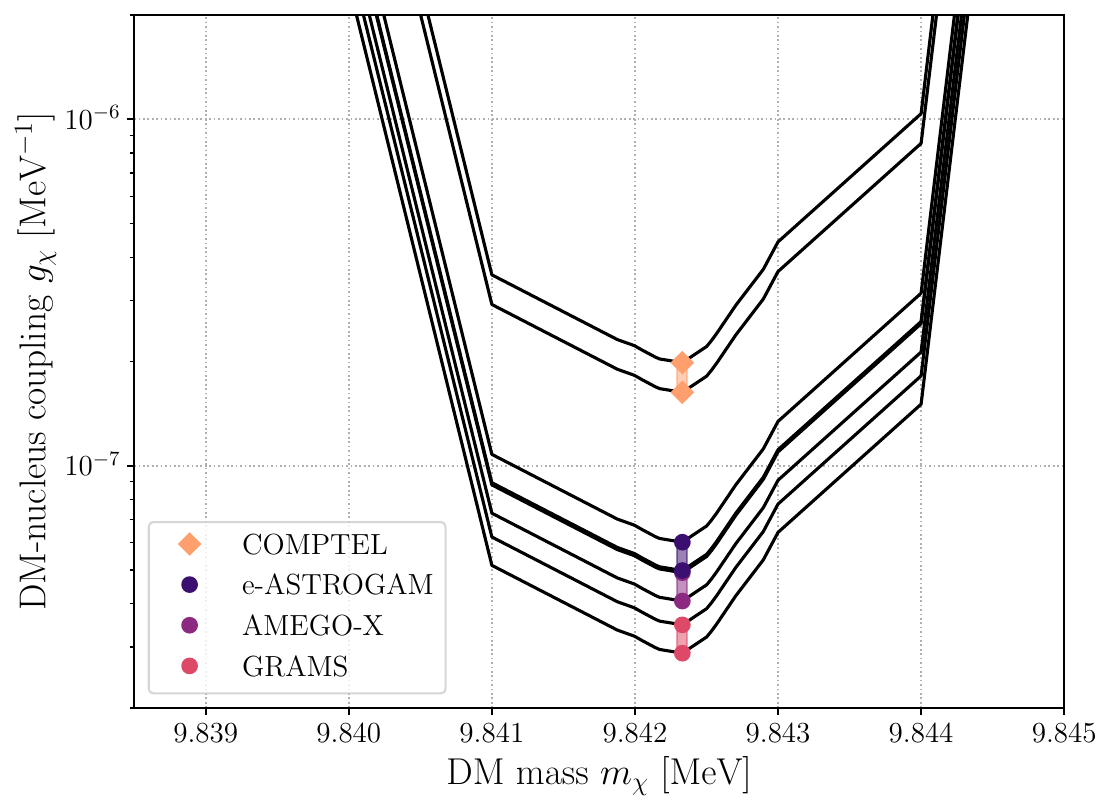}
    \caption{
    Upper limits on $g_\chi$ for DM absorption by $\ce{^{12}C}$ for a single excitation energy, $\Delta E~\approx~9.845$ MeV, for DM masses that give a non-zero differential flux.
    Two points of the same color represent the upper and lower uncertainties on $g_\chi$ for a given experiment, determined by the data presented in Ref.~\cite{Strong:1998ck}.
    For this excitation energy, COMPTEL limits are derived by integrating over the 3--10 MeV energy bin.
    For future experiments e-ASTROGAM, AMEGO-X, and GRAMS, these limits represent 68\% confidence level as the convolved spectrum is integrated over the range $E_\gamma \pm \epsilon E_\gamma$.
    The constraint points are drawn only at the lowest point of the DM mass curve, i.e. at the most constraining DM mass for this nuclear excitation energy, which corresponds to a baryon velocity of $v_n \approx 219~\text{km/s}$ from Eq.~\eqref{eq:m_chi_deltaFunction}.
    Note that AMEGO-X's upper limit point overlaps with the lower limit point of e-ASTROGAM, and is therefore not visible in the figure.}
    \label{fig:one_mass}
\end{figure}

\begin{figure*}[hbt!]
    \centering
    \includegraphics[width=1\textwidth]{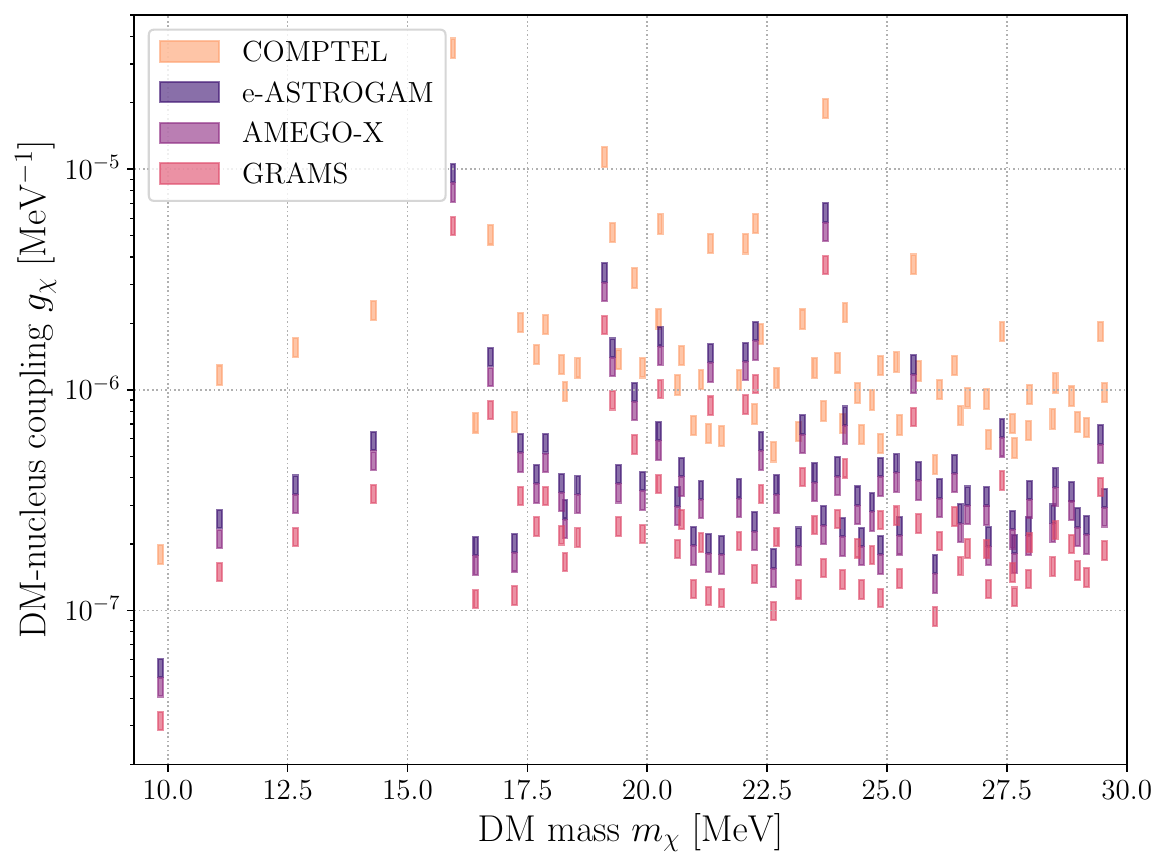}
    \caption{
    Upper limits on $g_\chi$ for DM absorption by $\ce{^{12}C}$ and $\ce{^{16}O}$, derived from existing COMPTEL data from Ref.~\cite{Strong:1998ck} (with assumed energy resolution for COMPTEL of $\epsilon = 2\%$), and projected constraints for future experiments e-ASTROGAM \cite{e-ASTROGAM:2016bph} (assumed energy resolution of $\epsilon = 3\%$), AMEGO-X \cite{Caputo:2022xpx} ($\epsilon = 2\%$), and GRAMS \cite{Aramaki:2019bpi} ($\epsilon = 1\%$). 
    Each shaded rectangle represents the possible values of $g_\chi$ for each $\ce{^{12}C}$ or $\ce{^{16}O}$ excitation energy shown in Fig.~\ref{fig:GT_strengths}, with upper and lower uncertainties determined by the data presented in Ref.~\cite{Strong:1998ck}.
    As in Fig.~\ref{fig:one_mass}, the constraints shown are from the most constraining DM mass for each nuclear excitation energy.
    The photon flux is found to be very sensitive to the value of $m_\chi$ such that the range of DM masses that will result in the excitation of the nucleus for a particular excitation energy $\Delta E$ is very narrow, as shown in Fig.~\ref{fig:one_mass} for a single $\ce{^{12}C}$ excitation energy.
    Here we omit the black line in Fig.~\ref{fig:one_mass} for clarity, which in this plot would appear as a vertical line.
    }
    \label{fig:all_limits}
\end{figure*}

We use COMPTEL data to calculate constraints on $g_\chi$ by insisting that the prediction of $\int dE\, E^2 \times \text{intensity}$ from the DM absorption process cannot exceed the observation within the appropriate energy bin.
Our results are shown in Fig.~\ref{fig:one_mass} for a single $\ce{^{12}C}$ excitation energy and in Fig.~\ref{fig:all_limits} for all $\ce{^{12}C}$ and $\ce{^{16}O}$ excitation energies up to 30 MeV.
The absorption process and thus the intensity spectra are very sensitive to $m_\chi$; therefore, results are sharply peaked near the nuclear de-excitation energy, approximating a delta function centered at 
\begin{equation}\label{eq:m_chi_deltaFunction}
    m_\chi \approx \Delta E - \frac{1}{2}m_n v_n^2
\end{equation}
with $v_n \approx 219 ~ \text{km/s}$.
The extreme sensitivity to the DM mass means that despite $\ce{^{12}C}$ and $\ce{^{16}O}$ excitation energies that appear to be very close to one another in Fig.~\ref{fig:GT_strengths}, we find that any DM mass that results in a nuclear excitation only excites a single energy transition.

\begin{table*}[th]
    \centering
    \begin{tabular}{|c|c|c|}
    \hline
        \textbf{Experiment} & \textbf{Energy range} & \textbf{Energy resolution} $\mathbf{\epsilon}$ \\ \hline
        COMPTEL \cite{1993ApJS...86..657S} & 0.75 - 30 MeV &  2-4\% (5-10\% FWHM) \\ \hline
        AMEGO-X \cite{Caputo:2022xpx} & 25 keV - 1 GeV & 2\% (5\% FWHM) at 1 MeV\\
        \hline
        e-ASTROGAM \cite{e-ASTROGAM:2016bph} & 30 keV - 200 MeV & 3.0\% at 1 MeV \\
        \hline
        GRAMS \cite{Aramaki:2019bpi} & 0.2 MeV - 100 MeV & 1\% at 2.5 MeV \\
        \hline
    \end{tabular}
    \caption{
    Energy resolutions for each experiment considered in our analysis.
    For COMPTEL, we use the lowest energy resolution of 2\%.
    For COMPTEL and AMEGO-X, we calculated $\epsilon$ from $\epsilon~=~\text{FWHM} / (2 \sqrt{2 \ln 2})$.
    The GRAMS energy resolution of 1\% is estimated from the nEXO experiment (see Ref.~\cite{Aramaki:2019bpi}).
    }
    \label{tab:expt_parameters}
\end{table*}

\subsection{Projections for future experiments}

As a proxy for the data observed by future experiments, we fit the COMPTEL data from Ref.~\cite{Strong:1998ck} to a smooth function over the full 1--30 MeV energy range for the upper and lower COMPTEL uncertainty bands to obtain
\begin{align}
\label{eq:future_data}
    & E^2 \times \text{intensity}\\
    &=
    \begin{cases}
    \nonumber
        (1.5\times10^{-3}) \ln\left( \frac{\text{E}}{\text{MeV}}\right) + (1.2\times10^{-2}) & \text{lower} \\
        (2.4\times10^{-3}) \ln\left( \frac{\text{E}}{\text{MeV}}\right) + (1.7\times10^{-2}) & \text{upper},
    \end{cases}
\end{align}
in units of MeV cm$^{-2}$ s$^{-1}$ sr$^{-1}$.
We assume idealized optimal energy bins, with each bin centered on the de-excitation photon energy $E_\gamma$ and with width $E_\gamma \pm\epsilon E_\gamma$. 
The differential flux in \Eq{eq:diff_flux} is integrated and averaged the over the same angular range used in Section~\ref{sec:constraints-COMPTEL}. 

We consider future experiments AMEGO-X, e-ASTROGAM, and GRAMS, and use the smallest $\epsilon$ provided by their respective references at the energies most relevant to our study, listed in Table~\ref{tab:expt_parameters}.
Following the same procedure as in Section~\ref{sec:constraints-COMPTEL}, we integrate over the energy range $E_\gamma \pm\epsilon E_\gamma$.
Our results are shown in Fig.~\ref{fig:one_mass} for a single $\ce{^{12}C}$ excitation energy and in Fig.~\ref{fig:all_limits} for all $\ce{^{12}C}$ and $\ce{^{16}O}$ excitation energies. 

Of the future experiments considered, e-ASTROGAM is the least constraining on $g_\chi$, having the largest expected energy resolution at $3\%$, while GRAMS with the best expected energy resolution of 1\% is the most constraining.
Comparing current COMPTEL data to projected sensitivities, limits on $g_\chi$ improve by almost an order of magnitude on average, but we note that these projections are optimistic because of the optimized energy binning that we use.
Even though COMPTEL's energy resolution of $2\%$ is the same as that of AMEGO-X, COMPTEL is much less constraining due to the large energy bins.

Current constraints on the on the $g_\chi$ vs $m_\chi$ parameter space exist in the context of axion-like particles (ALPs) coupled to nucleons and emitted from stars and from supernova (SN) cores~\cite{Lella:2023bfb}. 
The solar ALP flux search at the Sudbury Neutrino Observatory excludes $g_\chi \gtrsim 2\times10^{-8}~\text{MeV}^{-1}$ for $m_\chi \lesssim 2~\text{MeV}$, while the SN~1987A cooling constraint excludes $3 \times 10^{-13}$ MeV$^{-1} \lesssim g_\chi \lesssim 2 \times 10^{-9}$ MeV$^{-1}$ for DM masses $m_\chi \lesssim 100$ MeV, where we have related the DM-nucleon coupling constant $g_\chi$ to the DM-proton $p$ coupling $g_{ap}$ of Ref.~\cite{Lella:2023bfb} by $g_\chi = g_{ap} / 2 m_p$.
None of these constraints apply any restriction to the parameter space we are probing.

In this analysis, we have chosen the DM particle to be pseudo-scalar type.
If the DM particle is of any other type, the shape of the $g_{\chi}$ vs $m_{\chi}$ plot will remain the same, while the change in the sensitivity of the coupling strength $g_{\chi}$ will be $\lesssim 2$.
However, for each of the DM types, the allowed DM parameter space will be different; e.g., the vector-type DM parameter space is shown in Ref.~\cite{An:2014twa}.


\section{Conclusion}
\label{sec:conclusions}
In this work we consider the absorption of a pseudoscalar DM particle by a nucleus, resulting in nuclear excitation and the subsequent release of a photon.
For $\ce{^{12}C}$ and $\ce{^{16}O}$ present in the GC, we use the \texttt{BIGSTICK} large-scale shell-model code to calculate nuclear excitation levels from $\sim$ 10 to 30 MeV, energies detectable by gamma-ray experiments.
Using current COMPTEL data, we constrain the DM-nucleus coupling and provide projections for future experiments AMEGO-X, e-ASTROGAM, and GRAMS.

We find this nuclear excitation process to be extremely sensitive to the DM mass due to the narrow-width approximation of the absorption.
Relaxing this approximation would produce smoother limits, though still peaked at the same DM masses.

For the model considered, we estimate future experiments would improve constraints on the DM-nucleus coupling by almost an order of magnitude compared to current data.
We expect constraints on other types of DM particle to be within one order of magnitude of these results, since they differ only by the degree of freedom of the DM spin.
For example, the cross section of fermionic DM is $\sim 2$ times higher than that of scalar/pseudoscalar DM.

With new proposed telescopes such as AMEGO-X, e-ASTROGAM, GRAMS, and others such as COSI~\citep{Tomsick:2021wed} (expected to launch in 2027), and GECCO~\citep{Moiseev:2023zkv}, sensitivity in the MeV range is expected to improve significantly, bridging the gap between hard X-ray and soft gamma-ray astronomy. 
In particular, AMEGO-X, e-ASTROGAM, and GRAMS would provide unprecedented sensitivity to both MeV-scale line emissions and continuum emission, improving upon COMPTEL's sensitivity by several orders of magnitude. 
Such experiments would provide insight into the origin of the MeV continuum emission, in addition to improving sensitivity to indirect detection signals. 


\begin{acknowledgments}

We thank Regina Caputo for comments on the manuscript. 
K.B.\ and S.M.\ thank Can Kilic for useful conversations.
K.B.\ and S.M.\ acknowledge support from the National Science Foundation under Grant No.~PHY-2112884. L.S., B.D., A.E., and W.-C.H.\ acknowledge support from DOE Grant de-sc0010813 and by the Texas A\&M University System National Laboratories Office and Los Alamos National Laboratory.

\end{acknowledgments}

\clearpage

\bibliographystyle{apsrev4-2}

\bibliography{main.bib}

\begin{thebibliography}{57}%
\makeatletter
\providecommand \@ifxundefined [1]{%
 \@ifx{#1\undefined}
}%
\providecommand \@ifnum [1]{%
 \ifnum #1\expandafter \@firstoftwo
 \else \expandafter \@secondoftwo
 \fi
}%
\providecommand \@ifx [1]{%
 \ifx #1\expandafter \@firstoftwo
 \else \expandafter \@secondoftwo
 \fi
}%
\providecommand \natexlab [1]{#1}%
\providecommand \enquote  [1]{``#1''}%
\providecommand \bibnamefont  [1]{#1}%
\providecommand \bibfnamefont [1]{#1}%
\providecommand \citenamefont [1]{#1}%
\providecommand \href@noop [0]{\@secondoftwo}%
\providecommand \href [0]{\begingroup \@sanitize@url \@href}%
\providecommand \@href[1]{\@@startlink{#1}\@@href}%
\providecommand \@@href[1]{\endgroup#1\@@endlink}%
\providecommand \@sanitize@url [0]{\catcode `\\12\catcode `\$12\catcode
  `\&12\catcode `\#12\catcode `\^12\catcode `\_12\catcode `\%12\relax}%
\providecommand \@@startlink[1]{}%
\providecommand \@@endlink[0]{}%
\providecommand \url  [0]{\begingroup\@sanitize@url \@url }%
\providecommand \@url [1]{\endgroup\@href {#1}{\urlprefix }}%
\providecommand \urlprefix  [0]{URL }%
\providecommand \Eprint [0]{\href }%
\providecommand \doibase [0]{https://doi.org/}%
\providecommand \selectlanguage [0]{\@gobble}%
\providecommand \bibinfo  [0]{\@secondoftwo}%
\providecommand \bibfield  [0]{\@secondoftwo}%
\providecommand \translation [1]{[#1]}%
\providecommand \BibitemOpen [0]{}%
\providecommand \bibitemStop [0]{}%
\providecommand \bibitemNoStop [0]{.\EOS\space}%
\providecommand \EOS [0]{\spacefactor3000\relax}%
\providecommand \BibitemShut  [1]{\csname bibitem#1\endcsname}%
\let\auto@bib@innerbib\@empty
\bibitem [{\citenamefont {Kaplan}\ \emph {et~al.}(2009)\citenamefont {Kaplan},
  \citenamefont {Luty},\ and\ \citenamefont {Zurek}}]{Kaplan:2009ag}%
  \BibitemOpen
  \bibfield  {author} {\bibinfo {author} {\bibfnamefont {D.~E.}\ \bibnamefont
  {Kaplan}}, \bibinfo {author} {\bibfnamefont {M.~A.}\ \bibnamefont {Luty}},\
  and\ \bibinfo {author} {\bibfnamefont {K.~M.}\ \bibnamefont {Zurek}},\ }\href
  {https://doi.org/10.1103/PhysRevD.79.115016} {\bibfield  {journal} {\bibinfo
  {journal} {Phys. Rev. D}\ }\textbf {\bibinfo {volume} {79}},\ \bibinfo
  {pages} {115016} (\bibinfo {year} {2009})},\ \Eprint
  {https://arxiv.org/abs/0901.4117} {arXiv:0901.4117 [hep-ph]} \BibitemShut
  {NoStop}%
\bibitem [{\citenamefont {Boddy}\ \emph {et~al.}(2022)\citenamefont {Boddy}
  \emph {et~al.}}]{Boddy:2022knd}%
  \BibitemOpen
  \bibfield  {author} {\bibinfo {author} {\bibfnamefont {K.~K.}\ \bibnamefont
  {Boddy}} \emph {et~al.},\ }\href
  {https://doi.org/10.1016/j.jheap.2022.06.005} {\bibfield  {journal} {\bibinfo
   {journal} {JHEAp}\ }\textbf {\bibinfo {volume} {35}},\ \bibinfo {pages}
  {112} (\bibinfo {year} {2022})},\ \Eprint {https://arxiv.org/abs/2203.06380}
  {arXiv:2203.06380 [hep-ph]} \BibitemShut {NoStop}%
\bibitem [{\citenamefont {Vergados}\ \emph {et~al.}(2022)\citenamefont
  {Vergados}, \citenamefont {Divari},\ and\ \citenamefont
  {Ejiri}}]{Vergados:2021ejk}%
  \BibitemOpen
  \bibfield  {author} {\bibinfo {author} {\bibfnamefont {J.~D.}\ \bibnamefont
  {Vergados}}, \bibinfo {author} {\bibfnamefont {P.~C.}\ \bibnamefont
  {Divari}},\ and\ \bibinfo {author} {\bibfnamefont {H.}~\bibnamefont
  {Ejiri}},\ }\href {https://doi.org/10.1155/2022/7373365} {\bibfield
  {journal} {\bibinfo  {journal} {Adv. High Energy Phys.}\ }\textbf {\bibinfo
  {volume} {2022}},\ \bibinfo {pages} {7373365} (\bibinfo {year} {2022})},\
  \Eprint {https://arxiv.org/abs/2104.12213} {arXiv:2104.12213 [hep-ph]}
  \BibitemShut {NoStop}%
\bibitem [{\citenamefont {Pospelov}\ \emph {et~al.}(2008)\citenamefont
  {Pospelov}, \citenamefont {Ritz},\ and\ \citenamefont
  {Voloshin}}]{Pospelov:2008jk}%
  \BibitemOpen
  \bibfield  {author} {\bibinfo {author} {\bibfnamefont {M.}~\bibnamefont
  {Pospelov}}, \bibinfo {author} {\bibfnamefont {A.}~\bibnamefont {Ritz}},\
  and\ \bibinfo {author} {\bibfnamefont {M.~B.}\ \bibnamefont {Voloshin}},\
  }\href {https://doi.org/10.1103/PhysRevD.78.115012} {\bibfield  {journal}
  {\bibinfo  {journal} {Phys. Rev. D}\ }\textbf {\bibinfo {volume} {78}},\
  \bibinfo {pages} {115012} (\bibinfo {year} {2008})},\ \Eprint
  {https://arxiv.org/abs/0807.3279} {arXiv:0807.3279 [hep-ph]} \BibitemShut
  {NoStop}%
\bibitem [{\citenamefont {Hochberg}\ \emph {et~al.}(2017)\citenamefont
  {Hochberg}, \citenamefont {Lin},\ and\ \citenamefont
  {Zurek}}]{Hochberg:2016sqx}%
  \BibitemOpen
  \bibfield  {author} {\bibinfo {author} {\bibfnamefont {Y.}~\bibnamefont
  {Hochberg}}, \bibinfo {author} {\bibfnamefont {T.}~\bibnamefont {Lin}},\ and\
  \bibinfo {author} {\bibfnamefont {K.~M.}\ \bibnamefont {Zurek}},\ }\href
  {https://doi.org/10.1103/PhysRevD.95.023013} {\bibfield  {journal} {\bibinfo
  {journal} {Phys. Rev. D}\ }\textbf {\bibinfo {volume} {95}},\ \bibinfo
  {pages} {023013} (\bibinfo {year} {2017})},\ \Eprint
  {https://arxiv.org/abs/1608.01994} {arXiv:1608.01994 [hep-ph]} \BibitemShut
  {NoStop}%
\bibitem [{\citenamefont {Benhabiles-Mezhoud}\ \emph
  {et~al.}(2013)\citenamefont {Benhabiles-Mezhoud}, \citenamefont {Kiener},
  \citenamefont {Tatischeff},\ and\ \citenamefont
  {Strong}}]{Benhabiles-Mezhoud_2013}%
  \BibitemOpen
  \bibfield  {author} {\bibinfo {author} {\bibfnamefont {H.}~\bibnamefont
  {Benhabiles-Mezhoud}}, \bibinfo {author} {\bibfnamefont {J.}~\bibnamefont
  {Kiener}}, \bibinfo {author} {\bibfnamefont {V.}~\bibnamefont {Tatischeff}},\
  and\ \bibinfo {author} {\bibfnamefont {A.~W.}\ \bibnamefont {Strong}},\
  }\href {https://doi.org/10.1088/0004-637X/763/2/98} {\bibfield  {journal}
  {\bibinfo  {journal} {The Astrophysical Journal}\ }\textbf {\bibinfo {volume}
  {763}},\ \bibinfo {pages} {98} (\bibinfo {year} {2013})}\BibitemShut
  {NoStop}%
\bibitem [{\citenamefont {{De Angelis}}\ \emph {et~al.}(2021)\citenamefont {{De
  Angelis}} \emph {et~al.}}]{2021ExA....51.1225D}%
  \BibitemOpen
  \bibfield  {author} {\bibinfo {author} {\bibfnamefont {A.}~\bibnamefont {{De
  Angelis}}} \emph {et~al.},\ }\href
  {https://doi.org/10.1007/s10686-021-09706-y} {\bibfield  {journal} {\bibinfo
  {journal} {Experimental Astronomy}\ }\textbf {\bibinfo {volume} {51}},\
  \bibinfo {pages} {1225} (\bibinfo {year} {2021})},\ \Eprint
  {https://arxiv.org/abs/2102.02460} {arXiv:2102.02460 [astro-ph.IM]}
  \BibitemShut {NoStop}%
\bibitem [{\citenamefont {Manna}\ and\ \citenamefont
  {Desai}(2024)}]{Manna:2023vsv}%
  \BibitemOpen
  \bibfield  {author} {\bibinfo {author} {\bibfnamefont {S.}~\bibnamefont
  {Manna}}\ and\ \bibinfo {author} {\bibfnamefont {S.}~\bibnamefont {Desai}},\
  }\href {https://doi.org/10.1088/1475-7516/2024/01/017} {\bibfield  {journal}
  {\bibinfo  {journal} {JCAP}\ }\textbf {\bibinfo {volume} {01}},\ \bibinfo
  {pages} {017}},\ \Eprint {https://arxiv.org/abs/2310.07519} {arXiv:2310.07519
  [astro-ph.HE]} \BibitemShut {NoStop}%
\bibitem [{\citenamefont {{Diehl}}\ and\ \citenamefont
  {{Timmes}}(1998)}]{1998PASP..110..637D}%
  \BibitemOpen
  \bibfield  {author} {\bibinfo {author} {\bibfnamefont {R.}~\bibnamefont
  {{Diehl}}}\ and\ \bibinfo {author} {\bibfnamefont {F.~X.}\ \bibnamefont
  {{Timmes}}},\ }\href {https://doi.org/10.1086/316169} {\bibfield  {journal}
  {\bibinfo  {journal} {PASP}\ }\textbf {\bibinfo {volume} {110}},\ \bibinfo
  {pages} {637} (\bibinfo {year} {1998})}\BibitemShut {NoStop}%
\bibitem [{\citenamefont {{Iyudin}}\ \emph {et~al.}(2004)\citenamefont
  {{Iyudin}}, \citenamefont {{B{\"o}hringer}}, \citenamefont {{Dogiel}},\ and\
  \citenamefont {{Morfill}}}]{2004A&A...413..817I}%
  \BibitemOpen
  \bibfield  {author} {\bibinfo {author} {\bibfnamefont {A.~F.}\ \bibnamefont
  {{Iyudin}}}, \bibinfo {author} {\bibfnamefont {H.}~\bibnamefont
  {{B{\"o}hringer}}}, \bibinfo {author} {\bibfnamefont {V.}~\bibnamefont
  {{Dogiel}}},\ and\ \bibinfo {author} {\bibfnamefont {G.}~\bibnamefont
  {{Morfill}}},\ }\href {https://doi.org/10.1051/0004-6361:20031599} {\bibfield
   {journal} {\bibinfo  {journal} {AAP}\ }\textbf {\bibinfo {volume} {413}},\
  \bibinfo {pages} {817} (\bibinfo {year} {2004})}\BibitemShut {NoStop}%
\bibitem [{\citenamefont {Wang}\ \emph {et~al.}(2007)\citenamefont {Wang} \emph
  {et~al.}}]{Wang:2007sa}%
  \BibitemOpen
  \bibfield  {author} {\bibinfo {author} {\bibfnamefont {W.}~\bibnamefont
  {Wang}} \emph {et~al.},\ }\href {https://doi.org/10.1051/0004-6361:20066982}
  {\bibfield  {journal} {\bibinfo  {journal} {Astron. Astrophys.}\ }\textbf
  {\bibinfo {volume} {469}},\ \bibinfo {pages} {1005} (\bibinfo {year}
  {2007})},\ \Eprint {https://arxiv.org/abs/0704.3895} {arXiv:0704.3895
  [astro-ph]} \BibitemShut {NoStop}%
\bibitem [{\citenamefont {{Glover}}\ and\ \citenamefont
  {{Clark}}(2016)}]{2016MNRAS.456.3596G}%
  \BibitemOpen
  \bibfield  {author} {\bibinfo {author} {\bibfnamefont {S.~C.~O.}\
  \bibnamefont {{Glover}}}\ and\ \bibinfo {author} {\bibfnamefont {P.~C.}\
  \bibnamefont {{Clark}}},\ }\href {https://doi.org/10.1093/mnras/stv2863}
  {\bibfield  {journal} {\bibinfo  {journal} {Monthly Notices of the Royal
  Astronomical Society}\ }\textbf {\bibinfo {volume} {456}},\ \bibinfo {pages}
  {3596} (\bibinfo {year} {2016})},\ \Eprint {https://arxiv.org/abs/1509.01939}
  {arXiv:1509.01939 [astro-ph.GA]} \BibitemShut {NoStop}%
\bibitem [{\citenamefont {Johnson}\ \emph {et~al.}(2018)\citenamefont
  {Johnson}, \citenamefont {Ormand}, \citenamefont {McElvain},\ and\
  \citenamefont {Shan}}]{johnson2018bigstick}%
  \BibitemOpen
  \bibfield  {author} {\bibinfo {author} {\bibfnamefont {C.~W.}\ \bibnamefont
  {Johnson}}, \bibinfo {author} {\bibfnamefont {W.~E.}\ \bibnamefont {Ormand}},
  \bibinfo {author} {\bibfnamefont {K.~S.}\ \bibnamefont {McElvain}},\ and\
  \bibinfo {author} {\bibfnamefont {H.}~\bibnamefont {Shan}},\ }\href@noop {}
  {\bibinfo {title} {{BIGSTICK}: A flexible configuration-interaction
  shell-model code}} (\bibinfo {year} {2018}),\ \Eprint
  {https://arxiv.org/abs/1801.08432} {arXiv:1801.08432 [physics.comp-ph]}
  \BibitemShut {NoStop}%
\bibitem [{\citenamefont {Johnson}\ \emph {et~al.}(2013)\citenamefont
  {Johnson}, \citenamefont {Ormand},\ and\ \citenamefont
  {Krastev}}]{Johnson:2013bna}%
  \BibitemOpen
  \bibfield  {author} {\bibinfo {author} {\bibfnamefont {C.~W.}\ \bibnamefont
  {Johnson}}, \bibinfo {author} {\bibfnamefont {W.~E.}\ \bibnamefont
  {Ormand}},\ and\ \bibinfo {author} {\bibfnamefont {P.~G.}\ \bibnamefont
  {Krastev}},\ }\href {https://doi.org/10.1016/j.cpc.2013.07.022} {\bibfield
  {journal} {\bibinfo  {journal} {Comput. Phys. Commun.}\ }\textbf {\bibinfo
  {volume} {184}},\ \bibinfo {pages} {2761} (\bibinfo {year} {2013})},\ \Eprint
  {https://arxiv.org/abs/1303.0905} {arXiv:1303.0905 [nucl-th]} \BibitemShut
  {NoStop}%
\bibitem [{\citenamefont {Kelley}\ \emph {et~al.}(2017)\citenamefont {Kelley},
  \citenamefont {Purcell},\ and\ \citenamefont {Sheu}}]{KELLEY201771}%
  \BibitemOpen
  \bibfield  {author} {\bibinfo {author} {\bibfnamefont {J.}~\bibnamefont
  {Kelley}}, \bibinfo {author} {\bibfnamefont {J.}~\bibnamefont {Purcell}},\
  and\ \bibinfo {author} {\bibfnamefont {C.}~\bibnamefont {Sheu}},\ }\href
  {https://doi.org/https://doi.org/10.1016/j.nuclphysa.2017.07.015} {\bibfield
  {journal} {\bibinfo  {journal} {Nuclear Physics A}\ }\textbf {\bibinfo
  {volume} {968}},\ \bibinfo {pages} {71} (\bibinfo {year} {2017})}\BibitemShut
  {NoStop}%
\bibitem [{\citenamefont {Tilley}\ \emph {et~al.}(1993)\citenamefont {Tilley},
  \citenamefont {Weller},\ and\ \citenamefont {Cheves}}]{TILLEY19931}%
  \BibitemOpen
  \bibfield  {author} {\bibinfo {author} {\bibfnamefont {D.}~\bibnamefont
  {Tilley}}, \bibinfo {author} {\bibfnamefont {H.}~\bibnamefont {Weller}},\
  and\ \bibinfo {author} {\bibfnamefont {C.}~\bibnamefont {Cheves}},\ }\href
  {https://doi.org/https://doi.org/10.1016/0375-9474(93)90073-7} {\bibfield
  {journal} {\bibinfo  {journal} {Nuclear Physics A}\ }\textbf {\bibinfo
  {volume} {564}},\ \bibinfo {pages} {1} (\bibinfo {year} {1993})}\BibitemShut
  {NoStop}%
\bibitem [{\citenamefont {{Schoenfelder}}\ \emph {et~al.}(1993)\citenamefont
  {{Schoenfelder}} \emph {et~al.}}]{1993ApJS...86..657S}%
  \BibitemOpen
  \bibfield  {author} {\bibinfo {author} {\bibfnamefont {V.}~\bibnamefont
  {{Schoenfelder}}} \emph {et~al.},\ }\href {https://doi.org/10.1086/191794}
  {\bibfield  {journal} {\bibinfo  {journal} {Astrophysical Journal
  Supplement}\ }\textbf {\bibinfo {volume} {86}},\ \bibinfo {pages} {657}
  (\bibinfo {year} {1993})}\BibitemShut {NoStop}%
\bibitem [{\citenamefont {Schoenfelder}(2000)}]{Schoenfelder:2000bu}%
  \BibitemOpen
  \bibfield  {author} {\bibinfo {author} {\bibfnamefont {V.}~\bibnamefont
  {Schoenfelder}} (\bibinfo {collaboration} {COMPTEL}),\ }\href
  {https://doi.org/10.1051/aas:2000101} {\bibfield  {journal} {\bibinfo
  {journal} {Astron. Astrophys. Suppl. Ser.}\ }\textbf {\bibinfo {volume}
  {143}},\ \bibinfo {pages} {145} (\bibinfo {year} {2000})},\ \Eprint
  {https://arxiv.org/abs/astro-ph/0002366} {arXiv:astro-ph/0002366}
  \BibitemShut {NoStop}%
\bibitem [{\citenamefont {Fleischhack}(2023)}]{Fleischhack:2023ube}%
  \BibitemOpen
  \bibfield  {author} {\bibinfo {author} {\bibfnamefont {H.}~\bibnamefont
  {Fleischhack}} (\bibinfo {collaboration} {AMEGO-X}),\ }\href
  {https://doi.org/10.1088/1742-6596/2429/1/012023} {\bibfield  {journal}
  {\bibinfo  {journal} {J. Phys. Conf. Ser.}\ }\textbf {\bibinfo {volume}
  {2429}},\ \bibinfo {pages} {012023} (\bibinfo {year} {2023})}\BibitemShut
  {NoStop}%
\bibitem [{\citenamefont {Caputo}\ \emph {et~al.}(2022)\citenamefont {Caputo}
  \emph {et~al.}}]{Caputo:2022xpx}%
  \BibitemOpen
  \bibfield  {author} {\bibinfo {author} {\bibfnamefont {R.}~\bibnamefont
  {Caputo}} \emph {et~al.},\ }\href
  {https://doi.org/10.1117/1.JATIS.8.4.044003} {\bibfield  {journal} {\bibinfo
  {journal} {J. Astron. Telesc. Instrum. Syst.}\ }\textbf {\bibinfo {volume}
  {8}},\ \bibinfo {pages} {044003} (\bibinfo {year} {2022})},\ \Eprint
  {https://arxiv.org/abs/2208.04990} {arXiv:2208.04990 [astro-ph.IM]}
  \BibitemShut {NoStop}%
\bibitem [{\citenamefont {Tatischeff}\ \emph {et~al.}(2018)\citenamefont
  {Tatischeff} \emph {et~al.}}]{e-ASTROGAM:2018jlu}%
  \BibitemOpen
  \bibfield  {author} {\bibinfo {author} {\bibfnamefont {V.}~\bibnamefont
  {Tatischeff}} \emph {et~al.} (\bibinfo {collaboration} {e-ASTROGAM}),\ }\href
  {https://doi.org/10.1117/12.2315151} {\bibfield  {journal} {\bibinfo
  {journal} {Proc. SPIE Int. Soc. Opt. Eng.}\ }\textbf {\bibinfo {volume}
  {10699}},\ \bibinfo {pages} {106992J} (\bibinfo {year} {2018})},\ \Eprint
  {https://arxiv.org/abs/1805.06435} {arXiv:1805.06435 [astro-ph.HE]}
  \BibitemShut {NoStop}%
\bibitem [{\citenamefont {Aramaki}\ \emph {et~al.}(2021)\citenamefont {Aramaki}
  \emph {et~al.}}]{GRAMS:2021tax}%
  \BibitemOpen
  \bibfield  {author} {\bibinfo {author} {\bibfnamefont {T.}~\bibnamefont
  {Aramaki}} \emph {et~al.} (\bibinfo {collaboration} {GRAMS}),\ }\href
  {https://doi.org/10.22323/1.395.0653} {\bibfield  {journal} {\bibinfo
  {journal} {PoS}\ }\textbf {\bibinfo {volume} {ICRC2021}},\ \bibinfo {pages}
  {653} (\bibinfo {year} {2021})}\BibitemShut {NoStop}%
\bibitem [{\citenamefont {Dror}\ \emph
  {et~al.}(2020{\natexlab{a}})\citenamefont {Dror}, \citenamefont {Elor},\ and\
  \citenamefont {Mcgehee}}]{Dror:2019dib}%
  \BibitemOpen
  \bibfield  {author} {\bibinfo {author} {\bibfnamefont {J.~A.}\ \bibnamefont
  {Dror}}, \bibinfo {author} {\bibfnamefont {G.}~\bibnamefont {Elor}},\ and\
  \bibinfo {author} {\bibfnamefont {R.}~\bibnamefont {Mcgehee}},\ }\href
  {https://doi.org/10.1007/JHEP02(2020)134} {\bibfield  {journal} {\bibinfo
  {journal} {JHEP}\ }\textbf {\bibinfo {volume} {02}},\ \bibinfo {pages}
  {134}},\ \Eprint {https://arxiv.org/abs/1908.10861} {arXiv:1908.10861
  [hep-ph]} \BibitemShut {NoStop}%
\bibitem [{\citenamefont {Dror}\ \emph
  {et~al.}(2020{\natexlab{b}})\citenamefont {Dror}, \citenamefont {Elor},\ and\
  \citenamefont {Mcgehee}}]{Dror:2019onn}%
  \BibitemOpen
  \bibfield  {author} {\bibinfo {author} {\bibfnamefont {J.~A.}\ \bibnamefont
  {Dror}}, \bibinfo {author} {\bibfnamefont {G.}~\bibnamefont {Elor}},\ and\
  \bibinfo {author} {\bibfnamefont {R.}~\bibnamefont {Mcgehee}},\ }\href
  {https://doi.org/10.1103/PhysRevLett.124.181301} {\bibfield  {journal}
  {\bibinfo  {journal} {Phys. Rev. Lett.}\ }\textbf {\bibinfo {volume} {124}},\
  \bibinfo {pages} {18} (\bibinfo {year} {2020}{\natexlab{b}})},\ \Eprint
  {https://arxiv.org/abs/1905.12635} {arXiv:1905.12635 [hep-ph]} \BibitemShut
  {NoStop}%
\bibitem [{\citenamefont {Dent}\ \emph {et~al.}(2024)\citenamefont {Dent},
  \citenamefont {Dutta},\ and\ \citenamefont {Xu}}]{Dent:2024yje}%
  \BibitemOpen
  \bibfield  {author} {\bibinfo {author} {\bibfnamefont {J.~B.}\ \bibnamefont
  {Dent}}, \bibinfo {author} {\bibfnamefont {B.}~\bibnamefont {Dutta}},\ and\
  \bibinfo {author} {\bibfnamefont {T.}~\bibnamefont {Xu}},\ }\href@noop {} {\
  (\bibinfo {year} {2024})},\ \Eprint {https://arxiv.org/abs/2404.02956}
  {arXiv:2404.02956 [hep-ph]} \BibitemShut {NoStop}%
\bibitem [{\citenamefont {Agashe}\ \emph {et~al.}(2022)\citenamefont {Agashe},
  \citenamefont {Chang}, \citenamefont {Clark}, \citenamefont {Dutta},
  \citenamefont {Tsai},\ and\ \citenamefont {Xu}}]{Agashe:2022jgk}%
  \BibitemOpen
  \bibfield  {author} {\bibinfo {author} {\bibfnamefont {K.}~\bibnamefont
  {Agashe}}, \bibinfo {author} {\bibfnamefont {J.~H.}\ \bibnamefont {Chang}},
  \bibinfo {author} {\bibfnamefont {S.~J.}\ \bibnamefont {Clark}}, \bibinfo
  {author} {\bibfnamefont {B.}~\bibnamefont {Dutta}}, \bibinfo {author}
  {\bibfnamefont {Y.}~\bibnamefont {Tsai}},\ and\ \bibinfo {author}
  {\bibfnamefont {T.}~\bibnamefont {Xu}},\ }\href
  {https://doi.org/10.1103/PhysRevD.105.123009} {\bibfield  {journal} {\bibinfo
   {journal} {Phys. Rev. D}\ }\textbf {\bibinfo {volume} {105}},\ \bibinfo
  {pages} {123009} (\bibinfo {year} {2022})},\ \Eprint
  {https://arxiv.org/abs/2202.04653} {arXiv:2202.04653 [astro-ph.CO]}
  \BibitemShut {NoStop}%
\bibitem [{\citenamefont {Coogan}\ \emph {et~al.}(2021)\citenamefont {Coogan},
  \citenamefont {Morrison},\ and\ \citenamefont {Profumo}}]{Coogan:2020tuf}%
  \BibitemOpen
  \bibfield  {author} {\bibinfo {author} {\bibfnamefont {A.}~\bibnamefont
  {Coogan}}, \bibinfo {author} {\bibfnamefont {L.}~\bibnamefont {Morrison}},\
  and\ \bibinfo {author} {\bibfnamefont {S.}~\bibnamefont {Profumo}},\ }\href
  {https://doi.org/10.1103/PhysRevLett.126.171101} {\bibfield  {journal}
  {\bibinfo  {journal} {Phys. Rev. Lett.}\ }\textbf {\bibinfo {volume} {126}},\
  \bibinfo {pages} {171101} (\bibinfo {year} {2021})},\ \Eprint
  {https://arxiv.org/abs/2010.04797} {arXiv:2010.04797 [astro-ph.CO]}
  \BibitemShut {NoStop}%
\bibitem [{\citenamefont {Xie}\ \emph {et~al.}(2024)\citenamefont {Xie},
  \citenamefont {Liu}, \citenamefont {Liu}, \citenamefont {Cai},\ and\
  \citenamefont {Yang}}]{Xie:2024eug}%
  \BibitemOpen
  \bibfield  {author} {\bibinfo {author} {\bibfnamefont {Z.}~\bibnamefont
  {Xie}}, \bibinfo {author} {\bibfnamefont {B.}~\bibnamefont {Liu}}, \bibinfo
  {author} {\bibfnamefont {J.}~\bibnamefont {Liu}}, \bibinfo {author}
  {\bibfnamefont {Y.-F.}\ \bibnamefont {Cai}},\ and\ \bibinfo {author}
  {\bibfnamefont {R.}~\bibnamefont {Yang}},\ }\href
  {https://doi.org/10.1103/PhysRevD.109.043020} {\bibfield  {journal} {\bibinfo
   {journal} {Phys. Rev. D}\ }\textbf {\bibinfo {volume} {109}},\ \bibinfo
  {pages} {043020} (\bibinfo {year} {2024})},\ \Eprint
  {https://arxiv.org/abs/2401.06440} {arXiv:2401.06440 [astro-ph.HE]}
  \BibitemShut {NoStop}%
\bibitem [{\citenamefont {Cheek}\ \emph {et~al.}(2022)\citenamefont {Cheek},
  \citenamefont {Heurtier}, \citenamefont {Perez-Gonzalez},\ and\ \citenamefont
  {Turner}}]{Cheek:2021odj}%
  \BibitemOpen
  \bibfield  {author} {\bibinfo {author} {\bibfnamefont {A.}~\bibnamefont
  {Cheek}}, \bibinfo {author} {\bibfnamefont {L.}~\bibnamefont {Heurtier}},
  \bibinfo {author} {\bibfnamefont {Y.~F.}\ \bibnamefont {Perez-Gonzalez}},\
  and\ \bibinfo {author} {\bibfnamefont {J.}~\bibnamefont {Turner}},\ }\href
  {https://doi.org/10.1103/PhysRevD.105.015022} {\bibfield  {journal} {\bibinfo
   {journal} {Phys. Rev. D}\ }\textbf {\bibinfo {volume} {105}},\ \bibinfo
  {pages} {015022} (\bibinfo {year} {2022})},\ \Eprint
  {https://arxiv.org/abs/2107.00013} {arXiv:2107.00013 [hep-ph]} \BibitemShut
  {NoStop}%
\bibitem [{\citenamefont {Baer}\ \emph {et~al.}(2015)\citenamefont {Baer},
  \citenamefont {Choi}, \citenamefont {Kim},\ and\ \citenamefont
  {Roszkowski}}]{Baer:2014eja}%
  \BibitemOpen
  \bibfield  {author} {\bibinfo {author} {\bibfnamefont {H.}~\bibnamefont
  {Baer}}, \bibinfo {author} {\bibfnamefont {K.-Y.}\ \bibnamefont {Choi}},
  \bibinfo {author} {\bibfnamefont {J.~E.}\ \bibnamefont {Kim}},\ and\ \bibinfo
  {author} {\bibfnamefont {L.}~\bibnamefont {Roszkowski}},\ }\href
  {https://doi.org/10.1016/j.physrep.2014.10.002} {\bibfield  {journal}
  {\bibinfo  {journal} {Phys. Rept.}\ }\textbf {\bibinfo {volume} {555}},\
  \bibinfo {pages} {1} (\bibinfo {year} {2015})},\ \Eprint
  {https://arxiv.org/abs/1407.0017} {arXiv:1407.0017 [hep-ph]} \BibitemShut
  {NoStop}%
\bibitem [{\citenamefont {Di~Luzio}\ \emph {et~al.}(2020)\citenamefont
  {Di~Luzio}, \citenamefont {Giannotti}, \citenamefont {Nardi},\ and\
  \citenamefont {Visinelli}}]{DiLuzio:2020wdo}%
  \BibitemOpen
  \bibfield  {author} {\bibinfo {author} {\bibfnamefont {L.}~\bibnamefont
  {Di~Luzio}}, \bibinfo {author} {\bibfnamefont {M.}~\bibnamefont {Giannotti}},
  \bibinfo {author} {\bibfnamefont {E.}~\bibnamefont {Nardi}},\ and\ \bibinfo
  {author} {\bibfnamefont {L.}~\bibnamefont {Visinelli}},\ }\href
  {https://doi.org/10.1016/j.physrep.2020.06.002} {\bibfield  {journal}
  {\bibinfo  {journal} {Phys. Rept.}\ }\textbf {\bibinfo {volume} {870}},\
  \bibinfo {pages} {1} (\bibinfo {year} {2020})},\ \Eprint
  {https://arxiv.org/abs/2003.01100} {arXiv:2003.01100 [hep-ph]} \BibitemShut
  {NoStop}%
\bibitem [{\citenamefont {Groom}\ \emph {et~al.}(2000)\citenamefont {Groom}
  \emph {et~al.}}]{PDG2000}%
  \BibitemOpen
  \bibfield  {author} {\bibinfo {author} {\bibfnamefont {D.}~\bibnamefont
  {Groom}} \emph {et~al.},\ }\href {http://pdg.lbl.gov} {\bibfield  {journal}
  {\bibinfo  {journal} {The European Physical Journal}\ }\textbf {\bibinfo
  {volume} {C15}},\ \bibinfo {pages} {1} (\bibinfo {year} {2000})}\BibitemShut
  {NoStop}%
\bibitem [{\citenamefont {Dutta}\ \emph {et~al.}(2022)\citenamefont {Dutta},
  \citenamefont {Huang}, \citenamefont {Newstead},\ and\ \citenamefont
  {Pandey}}]{Dutta:2022tav}%
  \BibitemOpen
  \bibfield  {author} {\bibinfo {author} {\bibfnamefont {B.}~\bibnamefont
  {Dutta}}, \bibinfo {author} {\bibfnamefont {W.-C.}\ \bibnamefont {Huang}},
  \bibinfo {author} {\bibfnamefont {J.~L.}\ \bibnamefont {Newstead}},\ and\
  \bibinfo {author} {\bibfnamefont {V.}~\bibnamefont {Pandey}},\ }\href
  {https://doi.org/10.1103/PhysRevD.106.113006} {\bibfield  {journal} {\bibinfo
   {journal} {Phys. Rev. D}\ }\textbf {\bibinfo {volume} {106}},\ \bibinfo
  {pages} {113006} (\bibinfo {year} {2022})},\ \Eprint
  {https://arxiv.org/abs/2206.08590} {arXiv:2206.08590 [hep-ph]} \BibitemShut
  {NoStop}%
\bibitem [{\citenamefont {Warburton}\ and\ \citenamefont
  {Brown}(1992)}]{PhysRevC.46.923}%
  \BibitemOpen
  \bibfield  {author} {\bibinfo {author} {\bibfnamefont {E.~K.}\ \bibnamefont
  {Warburton}}\ and\ \bibinfo {author} {\bibfnamefont {B.~A.}\ \bibnamefont
  {Brown}},\ }\href {https://doi.org/10.1103/PhysRevC.46.923} {\bibfield
  {journal} {\bibinfo  {journal} {Phys. Rev. C}\ }\textbf {\bibinfo {volume}
  {46}},\ \bibinfo {pages} {923} (\bibinfo {year} {1992})}\BibitemShut
  {NoStop}%
\bibitem [{\citenamefont {Yuan}\ \emph {et~al.}(2012)\citenamefont {Yuan},
  \citenamefont {Suzuki}, \citenamefont {Otsuka}, \citenamefont {Xu},\ and\
  \citenamefont {Tsunoda}}]{Yuan:2012zz}%
  \BibitemOpen
  \bibfield  {author} {\bibinfo {author} {\bibfnamefont {C.}~\bibnamefont
  {Yuan}}, \bibinfo {author} {\bibfnamefont {T.}~\bibnamefont {Suzuki}},
  \bibinfo {author} {\bibfnamefont {T.}~\bibnamefont {Otsuka}}, \bibinfo
  {author} {\bibfnamefont {F.}~\bibnamefont {Xu}},\ and\ \bibinfo {author}
  {\bibfnamefont {N.}~\bibnamefont {Tsunoda}},\ }\href
  {https://doi.org/10.1103/PhysRevC.85.064324} {\bibfield  {journal} {\bibinfo
  {journal} {Phys. Rev. C}\ }\textbf {\bibinfo {volume} {85}},\ \bibinfo
  {pages} {064324} (\bibinfo {year} {2012})},\ \Eprint
  {https://arxiv.org/abs/1209.5587} {arXiv:1209.5587 [nucl-th]} \BibitemShut
  {NoStop}%
\bibitem [{NDS(2024)}]{NDS}%
  \BibitemOpen
  \href {https://www-nds.iaea.org/relnsd/vcharthtml/VChartHTML.html} {\bibinfo
  {title} {{IAEA} {N}uclear {D}ata {S}ervices. {L}ive chart of nuclides,
  nuclear structure and decay data}} (\bibinfo {year} {2024})\BibitemShut
  {NoStop}%
\bibitem [{\citenamefont {{Lin}}\ and\ \citenamefont
  {{Li}}(2019)}]{2019MNRAS.487.5679L}%
  \BibitemOpen
  \bibfield  {author} {\bibinfo {author} {\bibfnamefont {H.-N.}\ \bibnamefont
  {{Lin}}}\ and\ \bibinfo {author} {\bibfnamefont {X.}~\bibnamefont {{Li}}},\
  }\href {https://doi.org/10.1093/mnras/stz1698} {\bibfield  {journal}
  {\bibinfo  {journal} {Monthly Notices of the Royal Astronomical Society}\
  }\textbf {\bibinfo {volume} {487}},\ \bibinfo {pages} {5679} (\bibinfo {year}
  {2019})},\ \Eprint {https://arxiv.org/abs/1906.08419} {arXiv:1906.08419
  [astro-ph.GA]} \BibitemShut {NoStop}%
\bibitem [{\citenamefont {Porter}\ \emph {et~al.}(2017)\citenamefont {Porter},
  \citenamefont {Johannesson},\ and\ \citenamefont
  {Moskalenko}}]{Porter:2017vaa}%
  \BibitemOpen
  \bibfield  {author} {\bibinfo {author} {\bibfnamefont {T.~A.}\ \bibnamefont
  {Porter}}, \bibinfo {author} {\bibfnamefont {G.}~\bibnamefont
  {Johannesson}},\ and\ \bibinfo {author} {\bibfnamefont {I.~V.}\ \bibnamefont
  {Moskalenko}},\ }\href {https://doi.org/10.3847/1538-4357/aa844d} {\bibfield
  {journal} {\bibinfo  {journal} {Astrophys. J.}\ }\textbf {\bibinfo {volume}
  {846}},\ \bibinfo {pages} {67} (\bibinfo {year} {2017})},\ \Eprint
  {https://arxiv.org/abs/1708.00816} {arXiv:1708.00816 [astro-ph.HE]}
  \BibitemShut {NoStop}%
\bibitem [{\citenamefont {Porter}\ \emph {et~al.}(2022)\citenamefont {Porter},
  \citenamefont {Johannesson},\ and\ \citenamefont
  {Moskalenko}}]{Porter:2021tlr}%
  \BibitemOpen
  \bibfield  {author} {\bibinfo {author} {\bibfnamefont {T.~A.}\ \bibnamefont
  {Porter}}, \bibinfo {author} {\bibfnamefont {G.}~\bibnamefont
  {Johannesson}},\ and\ \bibinfo {author} {\bibfnamefont {I.~V.}\ \bibnamefont
  {Moskalenko}},\ }\href {https://doi.org/10.3847/1538-4365/ac80f6} {\bibfield
  {journal} {\bibinfo  {journal} {Astrophys. J. Supp.}\ }\textbf {\bibinfo
  {volume} {262}},\ \bibinfo {pages} {30} (\bibinfo {year} {2022})},\ \Eprint
  {https://arxiv.org/abs/2112.12745} {arXiv:2112.12745 [astro-ph.HE]}
  \BibitemShut {NoStop}%
\bibitem [{\citenamefont {J\'ohannesson}\ \emph {et~al.}(2018)\citenamefont
  {J\'ohannesson}, \citenamefont {Porter},\ and\ \citenamefont
  {Moskalenko}}]{Johannesson:2018bit}%
  \BibitemOpen
  \bibfield  {author} {\bibinfo {author} {\bibfnamefont {G.}~\bibnamefont
  {J\'ohannesson}}, \bibinfo {author} {\bibfnamefont {T.~A.}\ \bibnamefont
  {Porter}},\ and\ \bibinfo {author} {\bibfnamefont {I.~V.}\ \bibnamefont
  {Moskalenko}},\ }\href {https://doi.org/10.3847/1538-4357/aab26e} {\bibfield
  {journal} {\bibinfo  {journal} {Astrophys. J.}\ }\textbf {\bibinfo {volume}
  {856}},\ \bibinfo {pages} {45} (\bibinfo {year} {2018})},\ \Eprint
  {https://arxiv.org/abs/1802.08646} {arXiv:1802.08646 [astro-ph.HE]}
  \BibitemShut {NoStop}%
\bibitem [{\citenamefont {{Dame}}\ \emph {et~al.}(2001)\citenamefont {{Dame}},
  \citenamefont {{Hartmann}},\ and\ \citenamefont
  {{Thaddeus}}}]{2001ApJ...547..792D}%
  \BibitemOpen
  \bibfield  {author} {\bibinfo {author} {\bibfnamefont {T.~M.}\ \bibnamefont
  {{Dame}}}, \bibinfo {author} {\bibfnamefont {D.}~\bibnamefont {{Hartmann}}},\
  and\ \bibinfo {author} {\bibfnamefont {P.}~\bibnamefont {{Thaddeus}}},\
  }\href {https://doi.org/10.1086/318388} {\bibfield  {journal} {\bibinfo
  {journal} {The Astrophysical Journal}\ }\textbf {\bibinfo {volume} {547}},\
  \bibinfo {pages} {792} (\bibinfo {year} {2001})},\ \Eprint
  {https://arxiv.org/abs/astro-ph/0009217} {arXiv:astro-ph/0009217 [astro-ph]}
  \BibitemShut {NoStop}%
\bibitem [{\citenamefont {{Clemens}}(1985)}]{Clemens1985}%
  \BibitemOpen
  \bibfield  {author} {\bibinfo {author} {\bibfnamefont {D.~P.}\ \bibnamefont
  {{Clemens}}},\ }\href {https://doi.org/10.1086/163386} {\bibfield  {journal}
  {\bibinfo  {journal} {The Astrophysical Journal}\ }\textbf {\bibinfo {volume}
  {295}},\ \bibinfo {pages} {422} (\bibinfo {year} {1985})}\BibitemShut
  {NoStop}%
\bibitem [{\citenamefont {{Gillessen}}\ \emph {et~al.}(2009)\citenamefont
  {{Gillessen}}, \citenamefont {{Eisenhauer}}, \citenamefont {{Fritz}},
  \citenamefont {{Bartko}}, \citenamefont {{Dodds-Eden}}, \citenamefont
  {{Pfuhl}}, \citenamefont {{Ott}},\ and\ \citenamefont
  {{Genzel}}}]{2009ApJ...707L.114G}%
  \BibitemOpen
  \bibfield  {author} {\bibinfo {author} {\bibfnamefont {S.}~\bibnamefont
  {{Gillessen}}}, \bibinfo {author} {\bibfnamefont {F.}~\bibnamefont
  {{Eisenhauer}}}, \bibinfo {author} {\bibfnamefont {T.~K.}\ \bibnamefont
  {{Fritz}}}, \bibinfo {author} {\bibfnamefont {H.}~\bibnamefont {{Bartko}}},
  \bibinfo {author} {\bibfnamefont {K.}~\bibnamefont {{Dodds-Eden}}}, \bibinfo
  {author} {\bibfnamefont {O.}~\bibnamefont {{Pfuhl}}}, \bibinfo {author}
  {\bibfnamefont {T.}~\bibnamefont {{Ott}}},\ and\ \bibinfo {author}
  {\bibfnamefont {R.}~\bibnamefont {{Genzel}}},\ }\href
  {https://doi.org/10.1088/0004-637X/707/2/L114} {\bibfield  {journal}
  {\bibinfo  {journal} {The Astrophysical Journal Letters}\ }\textbf {\bibinfo
  {volume} {707}},\ \bibinfo {pages} {L114} (\bibinfo {year} {2009})},\ \Eprint
  {https://arxiv.org/abs/0910.3069} {arXiv:0910.3069 [astro-ph.GA]}
  \BibitemShut {NoStop}%
\bibitem [{\citenamefont {{McKeown}}\ \emph {et~al.}(2022)\citenamefont
  {{McKeown}}, \citenamefont {{Bullock}}, \citenamefont {{Mercado}},
  \citenamefont {{Hafen}}, \citenamefont {{Boylan-Kolchin}}, \citenamefont
  {{Wetzel}}, \citenamefont {{Necib}}, \citenamefont {{Hopkins}},\ and\
  \citenamefont {{Yu}}}]{2022MNRAS.513...55M}%
  \BibitemOpen
  \bibfield  {author} {\bibinfo {author} {\bibfnamefont {D.}~\bibnamefont
  {{McKeown}}}, \bibinfo {author} {\bibfnamefont {J.~S.}\ \bibnamefont
  {{Bullock}}}, \bibinfo {author} {\bibfnamefont {F.~J.}\ \bibnamefont
  {{Mercado}}}, \bibinfo {author} {\bibfnamefont {Z.}~\bibnamefont {{Hafen}}},
  \bibinfo {author} {\bibfnamefont {M.}~\bibnamefont {{Boylan-Kolchin}}},
  \bibinfo {author} {\bibfnamefont {A.}~\bibnamefont {{Wetzel}}}, \bibinfo
  {author} {\bibfnamefont {L.}~\bibnamefont {{Necib}}}, \bibinfo {author}
  {\bibfnamefont {P.~F.}\ \bibnamefont {{Hopkins}}},\ and\ \bibinfo {author}
  {\bibfnamefont {S.}~\bibnamefont {{Yu}}},\ }\href
  {https://doi.org/10.1093/mnras/stac966} {\bibfield  {journal} {\bibinfo
  {journal} {Monthly Notices of the Royal Astronomical Society}\ }\textbf
  {\bibinfo {volume} {513}},\ \bibinfo {pages} {55} (\bibinfo {year} {2022})},\
  \Eprint {https://arxiv.org/abs/2111.03076} {arXiv:2111.03076 [astro-ph.GA]}
  \BibitemShut {NoStop}%
\bibitem [{\citenamefont {{Marasco}}\ \emph {et~al.}(2017)\citenamefont
  {{Marasco}}, \citenamefont {{Fraternali}}, \citenamefont {{van der Hulst}},\
  and\ \citenamefont {{Oosterloo}}}]{2017A&A...607A.106M}%
  \BibitemOpen
  \bibfield  {author} {\bibinfo {author} {\bibfnamefont {A.}~\bibnamefont
  {{Marasco}}}, \bibinfo {author} {\bibfnamefont {F.}~\bibnamefont
  {{Fraternali}}}, \bibinfo {author} {\bibfnamefont {J.~M.}\ \bibnamefont {{van
  der Hulst}}},\ and\ \bibinfo {author} {\bibfnamefont {T.}~\bibnamefont
  {{Oosterloo}}},\ }\href {https://doi.org/10.1051/0004-6361/201731054}
  {\bibfield  {journal} {\bibinfo  {journal} {A\&A}\ }\textbf {\bibinfo
  {volume} {607}},\ \bibinfo {eid} {A106} (\bibinfo {year} {2017})},\ \Eprint
  {https://arxiv.org/abs/1707.00743} {arXiv:1707.00743 [astro-ph.GA]}
  \BibitemShut {NoStop}%
\bibitem [{\citenamefont {{Koppelman}}\ and\ \citenamefont
  {{Helmi}}(2021)}]{2021A&A...649A.136K}%
  \BibitemOpen
  \bibfield  {author} {\bibinfo {author} {\bibfnamefont {H.~H.}\ \bibnamefont
  {{Koppelman}}}\ and\ \bibinfo {author} {\bibfnamefont {A.}~\bibnamefont
  {{Helmi}}},\ }\href {https://doi.org/10.1051/0004-6361/202038777} {\bibfield
  {journal} {\bibinfo  {journal} {A\&A}\ }\textbf {\bibinfo {volume} {649}},\
  \bibinfo {eid} {A136} (\bibinfo {year} {2021})},\ \Eprint
  {https://arxiv.org/abs/2006.16283} {arXiv:2006.16283 [astro-ph.GA]}
  \BibitemShut {NoStop}%
\bibitem [{\citenamefont {{Monari}}\ \emph {et~al.}(2018)\citenamefont
  {{Monari}}, \citenamefont {{Famaey}}, \citenamefont {{Carrillo}},
  \citenamefont {{Piffl}}, \citenamefont {{Steinmetz}}, \citenamefont {{Wyse}},
  \citenamefont {{Anders}}, \citenamefont {{Chiappini}},\ and\ \citenamefont
  {{Jan{\ss}en}}}]{2018A&A...616L...9M}%
  \BibitemOpen
  \bibfield  {author} {\bibinfo {author} {\bibfnamefont {G.}~\bibnamefont
  {{Monari}}}, \bibinfo {author} {\bibfnamefont {B.}~\bibnamefont {{Famaey}}},
  \bibinfo {author} {\bibfnamefont {I.}~\bibnamefont {{Carrillo}}}, \bibinfo
  {author} {\bibfnamefont {T.}~\bibnamefont {{Piffl}}}, \bibinfo {author}
  {\bibfnamefont {M.}~\bibnamefont {{Steinmetz}}}, \bibinfo {author}
  {\bibfnamefont {R.~F.~G.}\ \bibnamefont {{Wyse}}}, \bibinfo {author}
  {\bibfnamefont {F.}~\bibnamefont {{Anders}}}, \bibinfo {author}
  {\bibfnamefont {C.}~\bibnamefont {{Chiappini}}},\ and\ \bibinfo {author}
  {\bibfnamefont {K.}~\bibnamefont {{Jan{\ss}en}}},\ }\href
  {https://doi.org/10.1051/0004-6361/201833748} {\bibfield  {journal} {\bibinfo
   {journal} {A\&A}\ }\textbf {\bibinfo {volume} {616}},\ \bibinfo {eid} {L9}
  (\bibinfo {year} {2018})},\ \Eprint {https://arxiv.org/abs/1807.04565}
  {arXiv:1807.04565 [astro-ph.GA]} \BibitemShut {NoStop}%
\bibitem [{\citenamefont {Roche}\ \emph {et~al.}(2024)\citenamefont {Roche},
  \citenamefont {Necib}, \citenamefont {Lin}, \citenamefont {Ou},\ and\
  \citenamefont {Nguyen}}]{Roche:2024gcl}%
  \BibitemOpen
  \bibfield  {author} {\bibinfo {author} {\bibfnamefont {C.}~\bibnamefont
  {Roche}}, \bibinfo {author} {\bibfnamefont {L.}~\bibnamefont {Necib}},
  \bibinfo {author} {\bibfnamefont {T.}~\bibnamefont {Lin}}, \bibinfo {author}
  {\bibfnamefont {X.}~\bibnamefont {Ou}},\ and\ \bibinfo {author}
  {\bibfnamefont {T.}~\bibnamefont {Nguyen}},\ }\href@noop {} {\  (\bibinfo
  {year} {2024})},\ \Eprint {https://arxiv.org/abs/2402.00108}
  {arXiv:2402.00108 [astro-ph.GA]} \BibitemShut {NoStop}%
\bibitem [{\citenamefont {Piffl}\ \emph {et~al.}(2014)\citenamefont {Piffl}
  \emph {et~al.}}]{Piffl:2013mla}%
  \BibitemOpen
  \bibfield  {author} {\bibinfo {author} {\bibfnamefont {T.}~\bibnamefont
  {Piffl}} \emph {et~al.},\ }\href
  {https://doi.org/10.1051/0004-6361/201322531} {\bibfield  {journal} {\bibinfo
   {journal} {Astron. Astrophys.}\ }\textbf {\bibinfo {volume} {562}},\
  \bibinfo {pages} {A91} (\bibinfo {year} {2014})},\ \Eprint
  {https://arxiv.org/abs/1309.4293} {arXiv:1309.4293 [astro-ph.GA]}
  \BibitemShut {NoStop}%
\bibitem [{\citenamefont {Perelstein}\ and\ \citenamefont
  {Shakya}(2010)}]{Perelstein:2010at}%
  \BibitemOpen
  \bibfield  {author} {\bibinfo {author} {\bibfnamefont {M.}~\bibnamefont
  {Perelstein}}\ and\ \bibinfo {author} {\bibfnamefont {B.}~\bibnamefont
  {Shakya}},\ }\href {https://doi.org/10.1088/1475-7516/2010/10/016} {\bibfield
   {journal} {\bibinfo  {journal} {JCAP}\ }\textbf {\bibinfo {volume} {10}},\
  \bibinfo {pages} {016}},\ \Eprint {https://arxiv.org/abs/1007.0018}
  {arXiv:1007.0018 [astro-ph.HE]} \BibitemShut {NoStop}%
\bibitem [{\citenamefont {Strong}\ \emph {et~al.}(1999)\citenamefont {Strong},
  \citenamefont {Bloemen}, \citenamefont {Diehl}, \citenamefont {Hermsen},\
  and\ \citenamefont {Schoenfelder}}]{Strong:1998ck}%
  \BibitemOpen
  \bibfield  {author} {\bibinfo {author} {\bibfnamefont {A.~W.}\ \bibnamefont
  {Strong}}, \bibinfo {author} {\bibfnamefont {H.}~\bibnamefont {Bloemen}},
  \bibinfo {author} {\bibfnamefont {R.}~\bibnamefont {Diehl}}, \bibinfo
  {author} {\bibfnamefont {W.}~\bibnamefont {Hermsen}},\ and\ \bibinfo {author}
  {\bibfnamefont {V.}~\bibnamefont {Schoenfelder}},\ }\href@noop {} {\bibfield
  {journal} {\bibinfo  {journal} {Astrophys. Lett. Commun.}\ }\textbf {\bibinfo
  {volume} {39}},\ \bibinfo {pages} {209} (\bibinfo {year} {1999})},\ \Eprint
  {https://arxiv.org/abs/astro-ph/9811211} {arXiv:astro-ph/9811211}
  \BibitemShut {NoStop}%
\bibitem [{\citenamefont {De~Angelis}\ \emph {et~al.}(2017)\citenamefont
  {De~Angelis} \emph {et~al.}}]{e-ASTROGAM:2016bph}%
  \BibitemOpen
  \bibfield  {author} {\bibinfo {author} {\bibfnamefont {A.}~\bibnamefont
  {De~Angelis}} \emph {et~al.} (\bibinfo {collaboration} {e-ASTROGAM}),\ }\href
  {https://doi.org/10.1007/s10686-017-9533-6} {\bibfield  {journal} {\bibinfo
  {journal} {Exper. Astron.}\ }\textbf {\bibinfo {volume} {44}},\ \bibinfo
  {pages} {25} (\bibinfo {year} {2017})},\ \Eprint
  {https://arxiv.org/abs/1611.02232} {arXiv:1611.02232 [astro-ph.HE]}
  \BibitemShut {NoStop}%
\bibitem [{\citenamefont {Aramaki}\ \emph {et~al.}(2020)\citenamefont
  {Aramaki}, \citenamefont {Hansson~Adrian}, \citenamefont {Karagiorgi},\ and\
  \citenamefont {Odaka}}]{Aramaki:2019bpi}%
  \BibitemOpen
  \bibfield  {author} {\bibinfo {author} {\bibfnamefont {T.}~\bibnamefont
  {Aramaki}}, \bibinfo {author} {\bibfnamefont {P.}~\bibnamefont
  {Hansson~Adrian}}, \bibinfo {author} {\bibfnamefont {G.}~\bibnamefont
  {Karagiorgi}},\ and\ \bibinfo {author} {\bibfnamefont {H.}~\bibnamefont
  {Odaka}},\ }\href {https://doi.org/10.1016/j.astropartphys.2019.07.002}
  {\bibfield  {journal} {\bibinfo  {journal} {Astropart. Phys.}\ }\textbf
  {\bibinfo {volume} {114}},\ \bibinfo {pages} {107} (\bibinfo {year}
  {2020})},\ \Eprint {https://arxiv.org/abs/1901.03430} {arXiv:1901.03430
  [astro-ph.HE]} \BibitemShut {NoStop}%
\bibitem [{\citenamefont {Lella}\ \emph {et~al.}(2024)\citenamefont {Lella},
  \citenamefont {Carenza}, \citenamefont {Co'}, \citenamefont {Lucente},
  \citenamefont {Giannotti}, \citenamefont {Mirizzi},\ and\ \citenamefont
  {Rauscher}}]{Lella:2023bfb}%
  \BibitemOpen
  \bibfield  {author} {\bibinfo {author} {\bibfnamefont {A.}~\bibnamefont
  {Lella}}, \bibinfo {author} {\bibfnamefont {P.}~\bibnamefont {Carenza}},
  \bibinfo {author} {\bibfnamefont {G.}~\bibnamefont {Co'}}, \bibinfo {author}
  {\bibfnamefont {G.}~\bibnamefont {Lucente}}, \bibinfo {author} {\bibfnamefont
  {M.}~\bibnamefont {Giannotti}}, \bibinfo {author} {\bibfnamefont
  {A.}~\bibnamefont {Mirizzi}},\ and\ \bibinfo {author} {\bibfnamefont
  {T.}~\bibnamefont {Rauscher}},\ }\href
  {https://doi.org/10.1103/PhysRevD.109.023001} {\bibfield  {journal} {\bibinfo
   {journal} {Phys. Rev. D}\ }\textbf {\bibinfo {volume} {109}},\ \bibinfo
  {pages} {023001} (\bibinfo {year} {2024})},\ \Eprint
  {https://arxiv.org/abs/2306.01048} {arXiv:2306.01048 [hep-ph]} \BibitemShut
  {NoStop}%
\bibitem [{\citenamefont {An}\ \emph {et~al.}(2015)\citenamefont {An},
  \citenamefont {Pospelov}, \citenamefont {Pradler},\ and\ \citenamefont
  {Ritz}}]{An:2014twa}%
  \BibitemOpen
  \bibfield  {author} {\bibinfo {author} {\bibfnamefont {H.}~\bibnamefont
  {An}}, \bibinfo {author} {\bibfnamefont {M.}~\bibnamefont {Pospelov}},
  \bibinfo {author} {\bibfnamefont {J.}~\bibnamefont {Pradler}},\ and\ \bibinfo
  {author} {\bibfnamefont {A.}~\bibnamefont {Ritz}},\ }\href
  {https://doi.org/10.1016/j.physletb.2015.06.018} {\bibfield  {journal}
  {\bibinfo  {journal} {Phys. Lett. B}\ }\textbf {\bibinfo {volume} {747}},\
  \bibinfo {pages} {331} (\bibinfo {year} {2015})},\ \Eprint
  {https://arxiv.org/abs/1412.8378} {arXiv:1412.8378 [hep-ph]} \BibitemShut
  {NoStop}%
\bibitem [{\citenamefont {Tomsick}(2021)}]{Tomsick:2021wed}%
  \BibitemOpen
  \bibfield  {author} {\bibinfo {author} {\bibfnamefont {J.~A.}\ \bibnamefont
  {Tomsick}} (\bibinfo {collaboration} {COSI}),\ }\href
  {https://doi.org/10.22323/1.395.0652} {\bibfield  {journal} {\bibinfo
  {journal} {PoS}\ }\textbf {\bibinfo {volume} {ICRC2021}},\ \bibinfo {pages}
  {652} (\bibinfo {year} {2021})},\ \Eprint {https://arxiv.org/abs/2109.10403}
  {arXiv:2109.10403 [astro-ph.IM]} \BibitemShut {NoStop}%
\bibitem [{\citenamefont {Moiseev}(2023)}]{Moiseev:2023zkv}%
  \BibitemOpen
  \bibfield  {author} {\bibinfo {author} {\bibfnamefont {A.}~\bibnamefont
  {Moiseev}} (\bibinfo {collaboration} {GECCO}),\ }\href
  {https://doi.org/10.22323/1.444.0702} {\bibfield  {journal} {\bibinfo
  {journal} {PoS}\ }\textbf {\bibinfo {volume} {ICRC2023}},\ \bibinfo {pages}
  {702} (\bibinfo {year} {2023})}\BibitemShut {NoStop}%
\end{thebibliography}%

\end{document}